\documentclass{aa}  
\usepackage{graphicx}
\usepackage{txfonts}
\begin{document}
\title{The evolution of the photometric properties of Local Group dwarf spheroidal galaxies}

	
\titlerunning{Photometric Properties of Local dSphs}

\author{F. Calura$^{\,1}$, G. A. Lanfranchi$^{\,2,3}$, 
F. Matteucci$^{\,1,3}$}
\institute{1 INAF - Osservatorio Astronomico di Trieste, via G. B. Tiepolo 11, 34131 Trieste, Italy \\
2 N\'ucleo de Astrof\'\i sica Te\'orica, CETEC, Universidade
Cruzeiro do Sul, R. Galv\~ao Bueno 868, Liberdade, 01506-000, S\~ao  Paulo, SP, Brazil \\
3 Dipartimento di Astronomia - Universit\'a di Trieste, Via G. B. Tiepolo 11, 34131 Trieste, Italy \\
}

\offprints{F. Calura, \email{fcalura@oats.inaf.it} }

\date{Received -; accepted -;}

\abstract
{}
{We investigate the present-day photometric properties of the dwarf spheroidal galaxies in the Local Group. 
From the analysis of their integrated colours, 
we consider a possible link between dwarf spheroidals and giant ellipticals. 
From the analysis of the $M_{V}$ vs (B-V) plot,  
we  search for a possible evolutionary link between dwarf spheroidal galaxies (dSphs) and dwarf irregular galaxies (dIrrs). }
{By means of chemical evolution models combined with a
spectro-photometric model, we study the evolution of six Local Group dwarf spheroidal
galaxies (Carina, Draco, Sagittarius, Sculptor, Sextans and Ursa
Minor). The chemical evolution models, which adopt up-to-date
nucleosynthesis from low and intermediate mass stars as well as  
nucleosynthesis and energetic feedback from
supernovae type Ia and II,  
reproduce several observational 
constraints of these galaxies, such as abundance ratios versus metallicity and 
the metallicity distributions. The 
proposed scenario for the evolution of these galaxies is characterised
by low star formation rates and high galactic wind efficiencies.}{ 
Such a scenario allows us to predict integrated colours and magnitudes which agree with
observations. 
Our results strongly suggest that the
first few Gyrs of evolution, when the star formation is most active, are 
crucial to define the luminosities, colours, and other
photometric properties as observed today. After the star formation epoch, 
 the galactic wind sweeps away a large fraction of the gas of each 
 galaxy, which then evolves passively. \\
Our results indicate that 
it is likely that at a certain stage of  their evolution, 
dSphs and dIrrs presented similar photometric properties. However, after that phase, they evolved 
along different paths, leading them to their currently disparate properties. }{}


\authorrunning{Calura, Lanfranchi, Matteucci}
\maketitle
\section{Introduction}

 The dwarf spheroidal galaxies (dSphs) are among the most common types
of galaxies in the universe. They are found normally in groups (C\^ot\'e et
al. 1997) and clusters (Phillips et al. 1998, Ferguson \& Sandage
1991). The ones found in the Local
Group have become an increasingly
important matter of study in the last few years due
to their proximity, which enables one to study in detail objects and
processes which were formally restricted to our own Galaxy. 

Several issues regarding the formation and evolution of the Local
Group dSphs can help in the attempt
to clarify the whole subject of galaxy formation. For example, how did the dSphs
form? When? What mechanism ruled their evolution? Are these
galaxies remnants of the building blocks from which larger galaxies
assembled? Is their evolution mainly affected by the environment or do internal
processes play a major role? Are these galaxies linked to other types
of dwarf galaxies in the context of any evolutionary scenario? In order
to answer these questions, one should try to understand not
only the present day properties of the dSphs, but also 
try to understand their past evolutionary history. One possible
procedure is to make use of models which, being based on
present day observational constraints, allow one to trace the past
evolution of the dSphs.
  
Originally, the local dSphs were
believed to be very old simple systems, similar to globular clusters 
(Shapley 1938). However, more recent deep photometric observations of
main sequence turn-off stars revealed also intermediate-age
populations, as well as a significant metallicity range, different from most globular clusters. In fact, analysis
of colour-magnitude diagrams (CMDs) suggested that these galaxies are
characterised by complex and different star formation (SF) histories 
(van den Bergh 1994; Hernandez, Gilmore, Valls-Gabaud 2000; Dolphin et
al. 2005). 
In almost all cases, the
mechanisms which trigger and control the SF are yet unknown and several
scenarios have been proposed. These scenarios should also explain the
complete lack of gas in these
galaxies, the low metallicities, the low values of [$\alpha$/Fe] 
relative to Galactic stars with the same [Fe/H]  
(Bonifacio et al. 2000; Shetrone, C\^ot\'e, Sargent 2001; Shetrone et
al. 2003; Tolstoy et al. 2003, Bonifacio et al. 2004;  Venn et
al. 2004; Sadakane et al. 2004; Geisler et al. 2005; Monaco et
al. 2005) and the metallicity distributions, including the large metallicity range (Koch et al. 2005, Bellazzini
et al. 2002). \\
Several attempts to model the properties of dSphs 
have been performed, generally following different approaches.
By means of high-resolution cosmological numerical simulations, 
Ricotti \& Gnedin (2005) and Kawata et al. (2006) studied dSphs in a cosmological framework. 
Their studies are useful to understand the link between the dSphs and the 
first galaxies, but cannot investigate in detail the chemical properties of dSphs and 
the relative roles of SNe Ia and II in the chemical enrichment and gas ejection processes. 
Marcolini et al. (2006), by means of a chemo-dynamical evolution model, 
studied the evolution of the interstellar medium (ISM) of the Draco dSph. In their picture 
no galactic wind develops and, in order to deplete the galaxy of its gas and to stop 
star formation, one must invoke an external mechanism, such as ram pressure stripping or tidal interactions. 
Fenner et al. (2006) and Ikuta $\&$ Arimoto (2002) studied the properties of dSphs by means 
of chemical evolution models including galactic winds. Their results indicate that 
a small fraction of the ISM is carried away by the SN-driven winds and confirm the need of external 
gas removing processes to reproduce the present-day gas fractions of dSphs (see also Mori \& Burkert 2000, Mayer et al. 2006). 
Lanfranchi $\&$ Matteucci
(2003, 2004, hereinafter LM03, LM04), alternatively, by means of a 
detailed chemical evolution 
model with galactic winds, were able to reproduce many observational constraints of six
local dSph galaxies (Carina, Draco, Sagittarius, Sculptor, Sextans,
Ursa Minor). The scenario proposed by these authors considered 
low efficiency star formation rates (SFR) derived from colour-magnitude diagrams together with 
intense galactic winds. \\
LM03,04 were able to reproduce not only the chemical properties,  
but also the lack of central neutral gas and the metallicity  distributions of the studied galaxies. In
particular, the  [$\alpha$/Fe]  vs [Fe/H] and neutron capture element ratios 
in these galaxies are well reproduced as well as the stellar metallicity
distribution (SMD).\\
The photometric properties of the local dSphs, however, are rarely
addressed in any of these models, even though they could help, not
only in constraining the formation and evolution of these galaxies,
but also in clarifying the subject of a possible evolutionary
connection between the gas poor dSph galaxies and gas rich dwarf
irregular galaxies (dIrrs). In such a scenario, a starburst in a dIrr
gives rise to a super wind which removes all the gas of the galaxy
(which could be removed also by ram pressure stripping or tidal
stripping) and halts the SF, giving rise to a dSph galaxy (Lin $\&$
Faber 1983; Dekel $\&$ Silk 1986;  van den
Bergh 1994; Papaderos et al. 1996; Davies $\&$ Phillipps 1988). There are, however, several
drawbacks in that scenario, from both the chemical 
and photometric point of view.  
First, the large scale distribution of dSphs is substantially different from that  of dIrrs. 
Most of the dSphs cluster around the two giant spirals of the Local Group (Grebel 1998), 
with very few exceptions. 
On the other hand, most of the dIrrs lie at large ($>500 kpc$) distances from the large galaxies (Mateo 1998). 
This is probably  linked to the morphology-density relation for dwarf galaxies 
and is interpreted as an evidence of environmental effects on galaxy evolution (Grebel 2001). In particular, it may 
be possible that their proximity to large galaxies had some effects on the evolution of dSphs.  \\
Futhermore, the dIrrs and dSphs show rather different observational properties. 
The integrated colour-magnitude diagram of the Local Group dwarf galaxies 
exhibits a clear distinction between the dSphs and the dIrrs, which occupy different regions (Mateo 1998). 
A few galaxies, namely transition type dwarfs, 
exhibit  intermediate photometric properies between those of dSphs and dIrrs. 
Also the luminosity/metallicity relation can be used to distinguish between
dIrrs and dSphs at the present epoch (Mateo 1998, Grebel et
al. 2003). At the same luminosity, the dIrrs tend to exhibit lower 
metallicities than the dSph galaxies. These two facts could be related
to different evolutionary histories for the two types of dwarfs and only
a study of their past history could help in clarifying this issue. 

In this work we show that a spectro-photometric code coupled
with a chemical evolution model allows us to investigate the past
evolution of the dSphs and impose constraints on the formation and
evolution of these galaxies and also to the possible connection with
other types of dwarf and giant galaxies. A chemical evolution code which reproduces
successfully several observational constraints can provide the
parameters required to predict the
evolution of the photometric properties of a sample of six local
dSph galaxies.

The paper is organized as follows: in Sect. 2, the chemical evolution
models which reproduce the chemical data of these galaxies and their
results are described. In Sect. 3 we describe the spectro-photometric
code. The results of  our models compared to observational data are
shown   in Sect. 4 and finally in Sect. 5 we draw some conclusions.

\section{Chemical Evolution Models} 

In order to provide the star formation rates, metallicities, gas mass, HI
mass and predictions of 
other parameters useful to the spectro-photometric code, we adopted
 the same chemical evolution models as described in our previous
 works. In particular, the models for Draco and Ursa Minor are taken
 from Lanfranchi $\&$ Matteucci (2007 - LM07), Carina and Sagittarius
 from Lanfranchi, Matteucci $\&$ Cescutti (2006b - LMC06b), Sculptor
 from Lanfranchi, Matteucci $\&$ Cescutti (2006a - LMC06a) and Sextans
 from LM04. 
The main parameters adopted for the model of each galaxy can be seen in 
Table 1. 
For more details on the models, we refer the reader to LM03, LM04. All these
 models already reproduced very well  the [$\alpha$/Fe] ratios, the
 [Ba/Fe] and [Eu/Fe] ratios, the stellar metallicity distributions and
 the present total mass and gas mass observed in these galaxies. In
 these models, we adopted a scenario in which the dSph galaxies form
 through a continuous and fast infall of pristine gas. 
We assume that the star formation histories (SFHs) of dSph galaxies consist of 
one or more episodes 
of different duration.  The main parameters in determining the star formation history 
consist in the burst duration $t_{B}$, the number of the bursts $n$ and the star formation efficiency $\nu$. 
The burst duration $t_{B}$ and the number of the bursts $n$ are 
provided by the observed 
colour-magnitude diagrams (CMDs), representing the most reliable constraints on the 
star formation history of each dSph (Dolphin 2002, Dolphin et al. 2005). 
%
These diagrams suggest a unique long ($t > 3 \;Gyr$) episode of SF in Draco,
Sextans, Sculptor, Ursa Minor and Sagittarius (Dolphin et al. 2005,
Carrera et al. 2002, Aparicio et al. 2001, Hernandez, Gilmore,
Valls-Gabaud 2000) and a few episodes (4) in Carina (Rizzi et
al. 2003). 
It is worth mentioning  that the models adopting the SFHs taken from observed CMDs 
(as described in  Table 2) reproduce very well the abundance ratios  
observed in these dSph galaxies. More important, the adopted 
SFHs
for Draco, Ursa Minor and Carina produced SMDs in very good agreement with
observations (LM07, LMC06b). In the case of Carina, LMC06b tested several
models
adopting in each one a different SFH (in particular the ones from Smecker-Hane et
al. 1996, Hurley-Keller, Mateo $\&$ Nemec 1998, Hernandez, Gilmore $\&$
Valls-Gabaud 2000, Dolphin 2002, and Rizzi et al. 2003)
and compared the predictions of such models with the [$\alpha$/Fe], [Ba/Fe],
[Eu/Fe]
and the SMD of Koch et al. (2006). The model with the Rizzi et al. (2003) SFH (the
same
adopted in this work) was the one which gave the best result.\\
\subsection{Basic ingredients} 
In table 2, we present the SFHs for the six dSph galaxies modelled in this work, along with the references 
to the SFHs inferred from studies of the observed CMDs, considered in this paper to compute the SFHs for the fiducial models. 

In the first column, we indicate the six dSphs. In the second, third and fourth  
columns, we show the number $n$, the times of occurrence $t_B$ and the durations $d$ of the star formation episodes, as inferred from the 
observed CMDs, respectively. Finally, in the last column we present 
the references for the observed CMDs used to determine the SFHs.\\
The galactic wind is actually one vital ingredient in these models since
it plays a crucial role in the evolution of the galaxies. When the
thermal energy of the gas equals or exceeds its binding
energy, a wind develops (see for example Matteucci $\&$ Tornamb\'e
1987). The thermal energy of the gas is controlled mainly by the
thermalization efficiencies of the supernovae explosions
($\eta_{SNII}$ for SNe II and $\eta_{SNIa}$ for SNeIa) and stellar
winds ($\eta_{SW}$), which control the fraction of the energy 
converted to thermal energy of the gas (see Bradamante
et al. 1998). The binding energy of the gas, on the other hand, is strongly
influenced by  assumptions concerning the presence and distribution
of dark matter (Matteucci 1992). A diffuse ($R_e/R_d$=0.1,  where
$R_e$ is the effective radius of the galaxy and $R_d$ is  the radius
of the dark matter core) but relatively massive  ($M_{dark}/M_{Lum}=10$) dark
halo has been assumed for each galaxy.  This particular configuration
allows the development of a galactic wind in these small systems
without destroying them. The galactic winds are intense (the rate of the wind is 4-13 times
the SFR) in the models for the dSph galaxies in order to reproduce the
decrease observed in the [$\alpha$/Fe] vs [Fe/H] plots, as well as several
features of the stellar metallicity distributions (LM07), and also to remove
a large fraction of the gas content of the galaxy. 
A possible
justification for intense galactic winds in dSph galaxies comes from the
fact that their potential well is not as deep as in the case of the
Dwarf Irregular galaxies and is more extended (Mateo 1998; Guzman et al 1998;
Graham \& Guzman 2003). In that sense, the gas
would be able to escape more easily from these systems. 
These models allow us to follow in detail the evolution of the
abundances of several chemical elements and isotopes, starting from  the matter
reprocessed by the stars and restored into the ISM 
by stellar winds and type II and Ia supernova explosions.
The type Ia SN progenitors are assumed to be C-O white dwarfs in binary 
systems according to the formalism originally developed by Greggio
$\&$ Renzini (1983) and Matteucci $\&$ Greggio (1986).
Type II SN explosions are assumed to originate from core collapse of single massive ($M > 8 M_{\odot}$) 
stars, leaving as remnants a neutron star or a black hole.

The main assumptions of the models are:

\begin{itemize}

\item
one zone with instantaneous and complete mixing of gas inside
this zone;

\item
no instantaneous recycling approximation, i.e. the stellar 
lifetimes are taken into account;

\item
the evolution of several chemical species (H, D, He, C, N, O, 
Mg, Si, S, Ca, Fe, Ba and Eu) is followed in detail;

\item
the nucleosynthesis prescriptions include the yields of 
Nomoto et al. (1997) for type Ia supernovae, Woosley $\&$
Weaver (1995) (with the corrections suggested by 
Fran\c cois et al., 2004) for massive stars ($M > 10 M_{\odot}$), 
van den Hoek $\&$ Groenewegen (1997) for intermediate mass stars 
(IMS) and for Ba and Eu the ones described in LMC06a and Cescutti et al. (2006). 

\end{itemize}

The basic equations of chemical evolution are the same as described 
in LM03 and LM04 (see also Tinsley 1980, Matteucci 1996), as are 
the prescriptions for the SFR $\psi(t)$, which follows a Schmidt law 
(Schmidt 1963): 
\begin{equation}
\psi(t) = \nu G^{k}(t)
\label{schmidt}
\end{equation}
where $G(t)$ is the gas fraction with respect to the total initial mass. 
$\nu$ is the SF efficiency, namely the SFR per unit mass of gas. For each dSph galaxy, the SF efficiency has been
determined in order to best reproduce the observed abundance ratios (LM04) and
the SMDs (LMC06b, LM07).\\ 
As in previous works (LM03, LM04), we assume $k=1$ and an 
initial mass function (IMF) from Salpeter (1955). 
The choice of the Salpeter IMF allows one to reproduce the 
abundances observed in dwarf galaxies (LM04, Recchi et al. 2002, LM07), as well as their present-day observational features (Calura \& Matteucci 2006). 
Moreover, as shown by Calura \& Matteucci (2004), the assumption of a Salpeter IMF allows one to account also for the local metal budget.\\
Our models include also infall and galactic winds, occurring at a rate $(\dot G_{i})$ and  $ (\dot{G_{i}})_{out}$, respectively. 
The rate of gas infall is defined as:
\begin{eqnarray}
(\dot G_{i})_{inf}\,=\,Ae^{-t/ \tau}
\end{eqnarray}
with $A$ being a suitable constant and $\tau$ the infall 
time-scale which is assumed to be 0.5 Gyr.

\begin{figure}
\centering
\vspace{0.001cm}
\includegraphics[height=19pc,width=16pc]{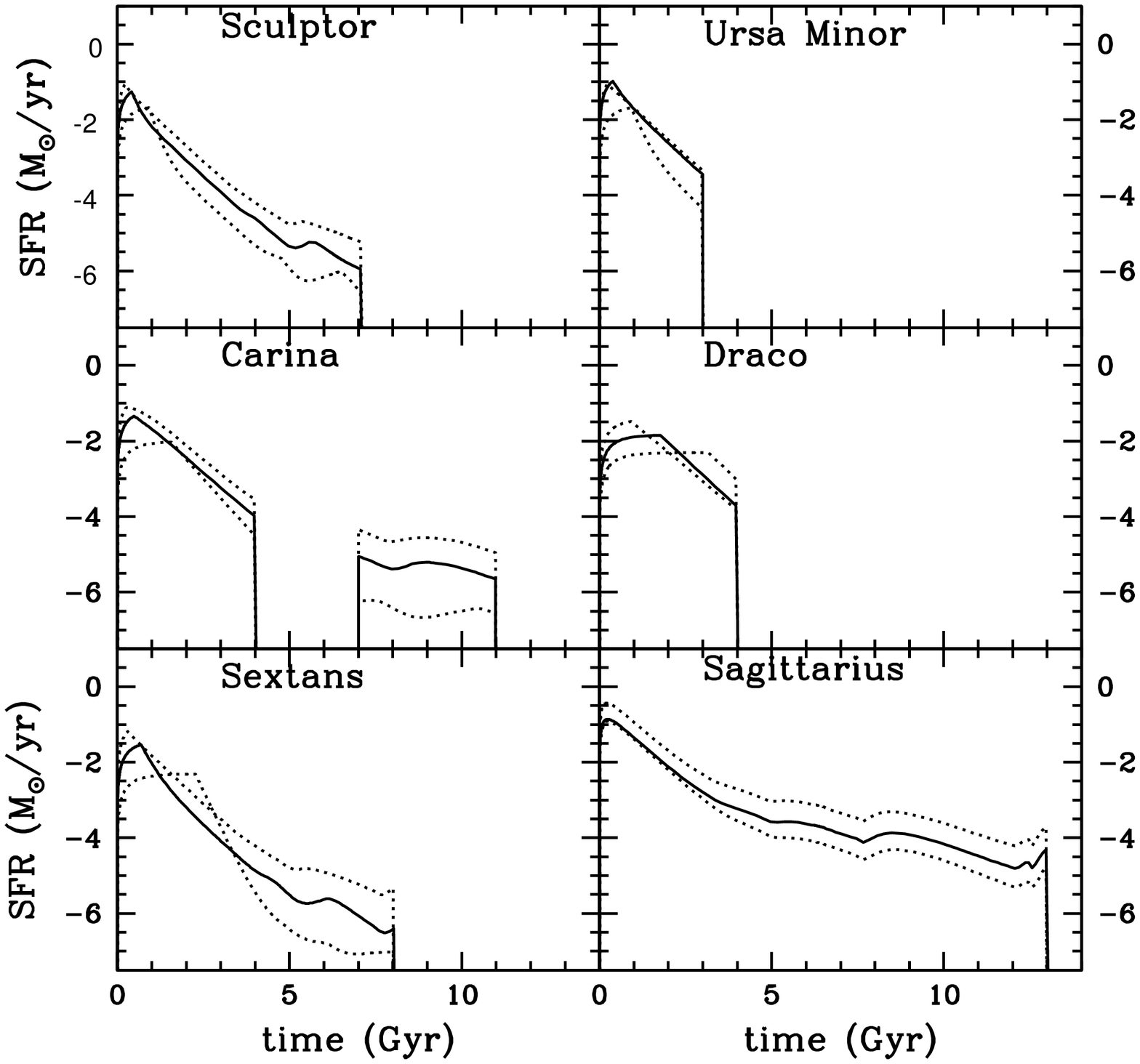}
\includegraphics[height=19pc,width=16pc]{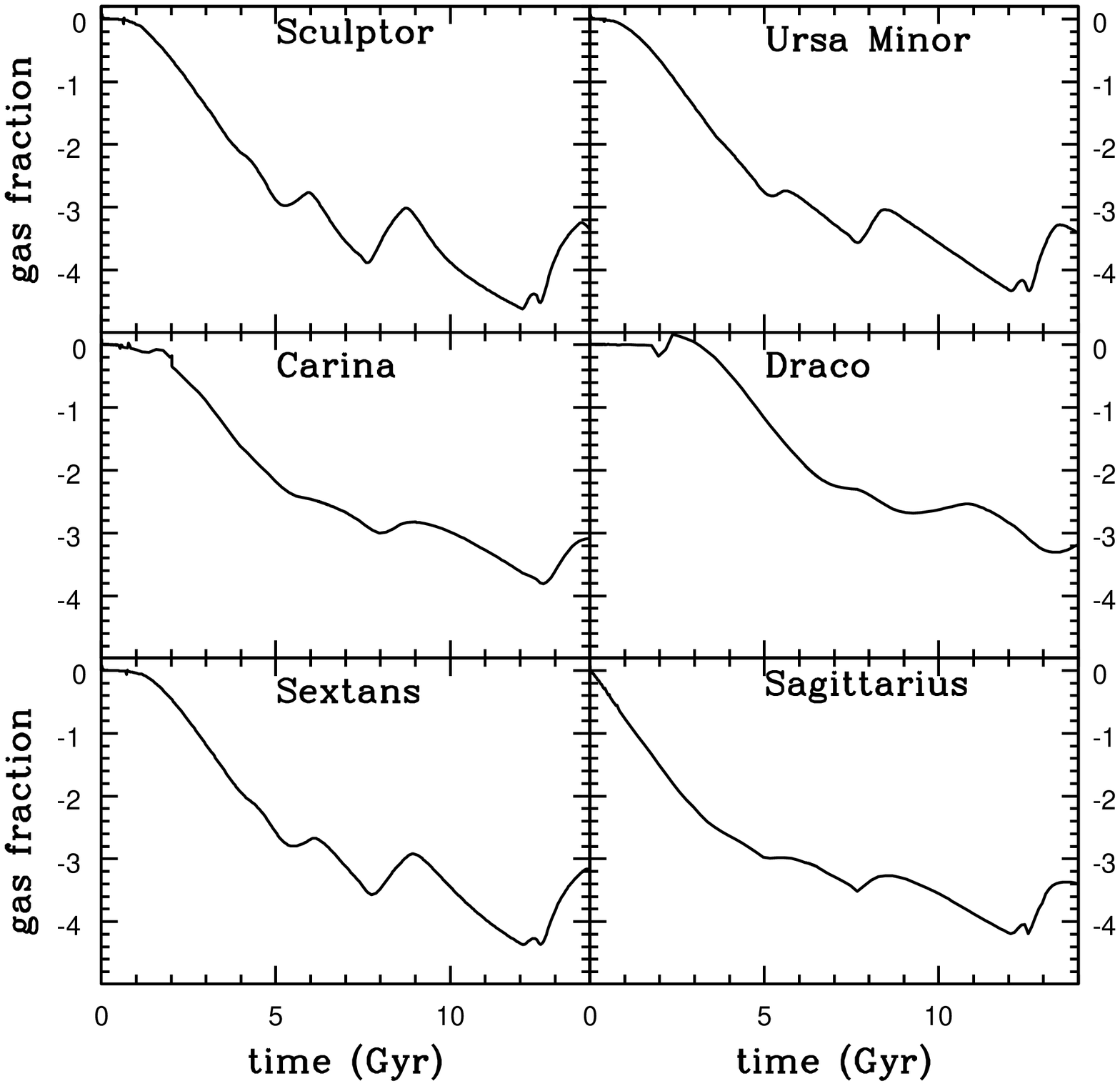}
\caption{ {\emph Upper panel:} evolution of the SFR as a function of time for the six dSphs. The solid lines represent the SFRs of the fiducial models. 
The dotted lines are star formation histories computed 
by assuming lower and upper limits for the star formation efficiencies, as indicated by the quantities $\nu_1$ and $\nu_2$ of Tab. 1, respectively. 
{\emph Lower panel:} evolution of the gas fractions for the fiducial models of the six dSphs (solid lines). }
\label{sfr_gas}
\end{figure}

The rate of gas loss via galactic winds for each element 
{\it i} is assumed to be proportional to the star formation 
rate at the time {\it t}:

\begin{eqnarray}
(\dot{G_{i}})_{out}\,=\,w_{i} \, \psi(t)
\end{eqnarray}
where $w_{i}$ is the wind efficiency parameter. \\ 
In the left panel of Fig.~\ref{sfr_gas}, we show the evolution of the star formation rates for the six dSphs considered in this work. 
For each galaxy, we show the fiducial star formation rates and the star formation rates computed 
by assuming lower and upper limits for the star formation efficiencies as in LM04. 
We will refer to these latter models as minimal and maximal model, respectively. 
The minimal and maximal models give an idea of the scatter in the SF 
efficiency parameter and of how wide its range is, within which one can still obtain 
a reasonably good fit to the abundance pattern observed in dSphs (LM04). \\
In some cases, it appears that the minimal and maximal lines may
reverse. This effect is related to the fact that, in each model, the time when the galactic wind develops 
depends on the assumed SF efficiency. 
In Fig. ~\ref{sfr_gas}, 
the onset of the wind corresponds to the change in slope of each curve. 
In general, the lower the SF efficiency, the later the galactic wind develops.  
For this reason, the lower the SF efficiency, 
the longer is the period during which the star formation remains high. 
This effect is particularly strong for the Draco and Sextans models, 
since these are the models presenting the widest range of values for $\nu$. \\
The lower and upper limits for the SF efficiency are indicated in Tab. 1 as $\nu_1$ and $\nu_2$, respectively. 
These quantities will be used later in Sect.s 4.4 - 4.5   to derive error bars for the present day colours of dSphs. 
Unless otherwise stated, all the calculations will be performed by using the fiducial star formation rates, 
computed using the SF efficiencies $\nu$ reported in Tab. 1.\\ 
In Fig. ~\ref{sfr_gas}, right panel, we plot the evolution of the gas fractions 
(defined as the ratio between the gas mass and the total baryonic mass, i.e. gas plus stars) for the six fiducial dSph models. 
As the gas is transformed into stars and removed by the winds, the gas fractions 
of the dSphs decrease by several orders of magnitude in few Gyrs. 
As shown in Eq.~\ref{schmidt}, at any time, the SFR depends on the amount of available gas present in the galaxy. 
For this reason, the star formation continues even after the onset of  the wind, but at a considerably lower rate. 
In the few last Gyrs of evolution, the oscillations in the gas content are due to injection of gas into the ISM from dying stars and to gas removal via 
winds and star formation. \\
Fenner et al. (2006) and Ikuta \& Arimoto (2002)  suggested that galactic winds do not play a major role in removing the gas from dSphs. 
In order to explain the lack of gas in dSph galaxies, they invoked a complementary mechanism, such as tidal  stripping or ram pressure. 
However, by means of closed-box models, Ikuta \& Arimoto (2002) are not  able to reproduce the lowest values of $[\alpha/Fe]$  
observed in dSphs without invoking a very steep IMF, whereas our models can reproduce those values very well with a normal Salpeter-like IMF
(see also  Geisler et al. 2007). 
Besides that, the works by Fenner et al. (2006) and Ikuta \& Arimoto (2002) do not 
compare their predictions with the stellar metallicity distributions (SMDs)  observed in dSph galaxies. On the other hand, 
Lanfranchi \& Matteucci (2007) consider the constraints provided by SMDs of local dSph galaxies, arguing that these diagrams 
can be reproduced only if strong galactic winds are taken into account. 

Recent evidence (Tolstoy et al. 2004) is finding that the Sculptor 
dSph has at least two distinct populations of stars, with different metallicity, 
spatial distributions, and kinematics.\\ 
These two different stellar populations 
can be explained as the consequence of a monolithic collapse  of gas, 
with more metal rich stars forming in the central regions  of the galaxy, and metal poor stars 
in the outskirts. 
Since our model is one-zone, we cannot distinguish between these two populations. 
However, we are able to reproduce quite well the overall chemical features for the dSphs. 
Furthermore, a photometric study of the spatial distribution of the stellar populations is beyond the aims of this paper, 
since here we are interested in the photometric properties of each dSph as a whole and 
not as a function of radius.

\renewcommand{\baselinestretch}{1.0}
\begin{table*}
\centering
\caption{Main parameters for the models of the six  dSph galaxies studied in this paper. 
$M_{tot}^{initial}$ (column 2) 
is the baryonic initial mass of the galaxy, in $M_{\odot}$. $w_{i}$ (column 3 ) 
is the wind efficiency. $\nu$ (column 4) is the star-formation 
efficiency,  a quantity expressing  
the intensity of each star formation event and is expressed in $Gyr^{-1}$.
Throughout this paper, $\nu$ is used to compute the \emph{fiducial} SFRs. 
$\nu_1$ and  $\nu_2$ (column 5) are the lower and uper limits on the SF efficiency, as shown in LM04. 
$\nu_1$ and  $\nu_2$ are used in Sect.s 4.4 and 4.5 to provide the errors associated with the theoretical 
predictions of the present-day 
colours of dSphs. Finally, the  
IMF (column 6) is the stellar initial mass function. } 
\begin{tabular}{lccccc}
\\[-2.0ex] 
\hline
\\[-2.5ex]
\noalign{\smallskip}  
galaxy &$M_{tot}^{initial} (M_{\odot})$ & $w_{i}$ & $\nu(Gyr^{-1})$ & $\nu_1, \nu_2 (Gyr^{-1})$ &$IMF$\\ 
\noalign{\smallskip}  
\hline
\\[-1.0ex]
Sextans  &$5\cdot10^{8}$&     9    & 0.08     &  0.01-0.3   &   Salpeter\\
Sculptor &$5\cdot10^{8}$&     13   & 0.2      &  0.05-0.5   &   Salpeter\\
Sagittarius &$5\cdot10^{8}$&  9    & 3.0   &  1.0-5.0    &   Salpeter\\
Draco  &$5\cdot10^{8}$ &      4    & 0.03       &  0.005-0.1  &   Salpeter\\
Ursa Minor &$5\cdot10^{8}$&  10    & 0.2    &  0.05-0.5   &   Salpeter\\
Carina &$5\cdot10^{8}$  &    5      & 0.15       &  0.02-0.4   &   Salpeter\\
\hline
\hline
\end{tabular}
\flushleft
\end{table*}

\begin{table*} 
\begin{center}
\caption[]{The adopted star formation histories for the dSph galaxies studied in this paper. In the first column, we list the dSphs. 
In the second, third and fourth column, 
$n$, $t_{B}$ and $d$ 
are the number, time of occurrence and duration of the SF 
episodes, respectively. In the last column, we present the references for the chosen SFHs. 
}
\begin{tabular}{c|ccc|c}  
\hline\hline\noalign{\smallskip}  
&\multicolumn{2}{c}{SFHs} & &  References\\
\noalign{\smallskip}  
\hline
& n &$t_B$ (Gyr) &$d$(Gyr) \\
\hline
Sextans      & 1  &0        &8               &LM04\\
Sculptor     & 1  &0        &7               &Dolphin 2002\\
Sagittarius  & 1  &0        &13              &Dolphin 2002\\
Draco        & 1  &0        &4               &Aparicio et al. 2001, \\
             &    &         &                &Dolphin et al. 2005\\
Ursa Minor   & 1  &0        &3               &Carrera et al. 2002, \\
             &    &         &                &Dolphin et al. 2005\\
Carina       & 4  &0/2/7/9  &2/2/2/2         &Rizzi et al. 2003\\
\hline\hline
\end{tabular}
\end{center}
\end{table*} 

\section{The spectro-photometric code}
To calculate galaxy colours and magnitudes, we use the photometric code by Bruzual \& Charlot 
(2003, hereinafter BC). This code allows one to compute the time evolution of the spectra of stellar 
populations in the metallicity range $0.0001 \le Z \le 0.05$, across the whole 
wavelength range from 3200  ${\rm \AA}$  to 9500  ${\rm \AA}$. 
We have implemented the BC code by taking into account the 
evolution of metallicity in galaxies (Calura, Matteucci \& Menci 2004). 
To model the spectral evolution of the 6 dSph galaxies 
studied in this paper, we have used simple stellar population (SSP) models calculated 
for a Salpeter (1955) IMF, with lower 
and upper cutoffs $m_{L}=0.1 M_{\odot}$ and  $m_{L}=100  M_{\odot}$, respectively. 
Dust extinction is also properly taken into account.  
We adopt the ``screen'' dust geometric distribution which, according to UV and optical 
observations of local starburst galaxies, is to be considered favored over the ``slab'' model 
(Calzetti et al. 1994). 
The absorbed flux $I_{a}(\lambda)$ of a stellar population behind a screen of dust is given by (Calzetti 2001):
\begin{equation}
I_{a}(\lambda)=I_{I}(\lambda) \, \exp{(-\tau(\lambda))}
\end{equation}
where $I_{I}(\lambda)$ represents the intrinsic, 
unobscured flux at the wavelength 
$\lambda$.

We assume that the optical depth $\tau(\lambda)$ is proportional to the gas surface density and to the metallicity 
$Z$, according to:
\begin{equation}
\tau(\lambda)= C \, \frac{k(\lambda) \, M_{gas} \, Z}{\pi\,r_d^2}  
\label{odepth}
\end{equation}
where $k(\lambda)$ is the extinction curve. 
In this paper, we adopt the 
extinction curve suggested by Calzetti (1997). 
$M_{gas}$ is the total gas mass and $r_d$ 
is the radius of the galaxy. We assume that, in each dSph galaxy, the neutral gas 
follows a homogeneous, spherical distribution within a radius $r_d=0.5$ kpc. 
This value is of the same order as the estimated tidal radii of some of these systems 
(Irwin \& Hatzidimitriou 1995, Gallagher et al. 2003).  
%
%
The constant C in equation~\ref{odepth} is chosen in order to reproduce the Milky Way average  
optical depth in the V band, $\tau_{V}=0.8$ (Calzetti 2001), given a gas surface density of $\sim 10 M_{\odot} pc^{-2}$ and the 
solar metallicity (Calura et al. 2004).   

\section{Results} 

\begin{figure}
\centering
\vspace{0.001cm}
\includegraphics[height=18pc,width=18pc]{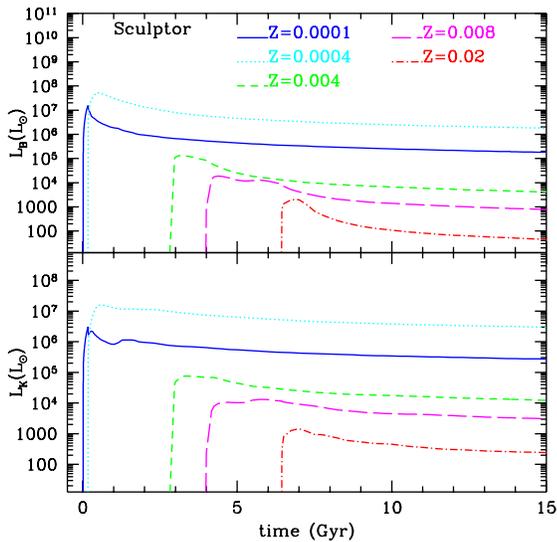}
\caption{ Predicted B- (upper panel) and K- (lower panel) band luminosities 
for the different stellar populations of the Sculptor dSph formed
at various metallicities as a function of time. }
\label{sculptor}
\end{figure}

\begin{figure}
\centering
\vspace{0.001cm}
\includegraphics[height=18pc,width=18pc]{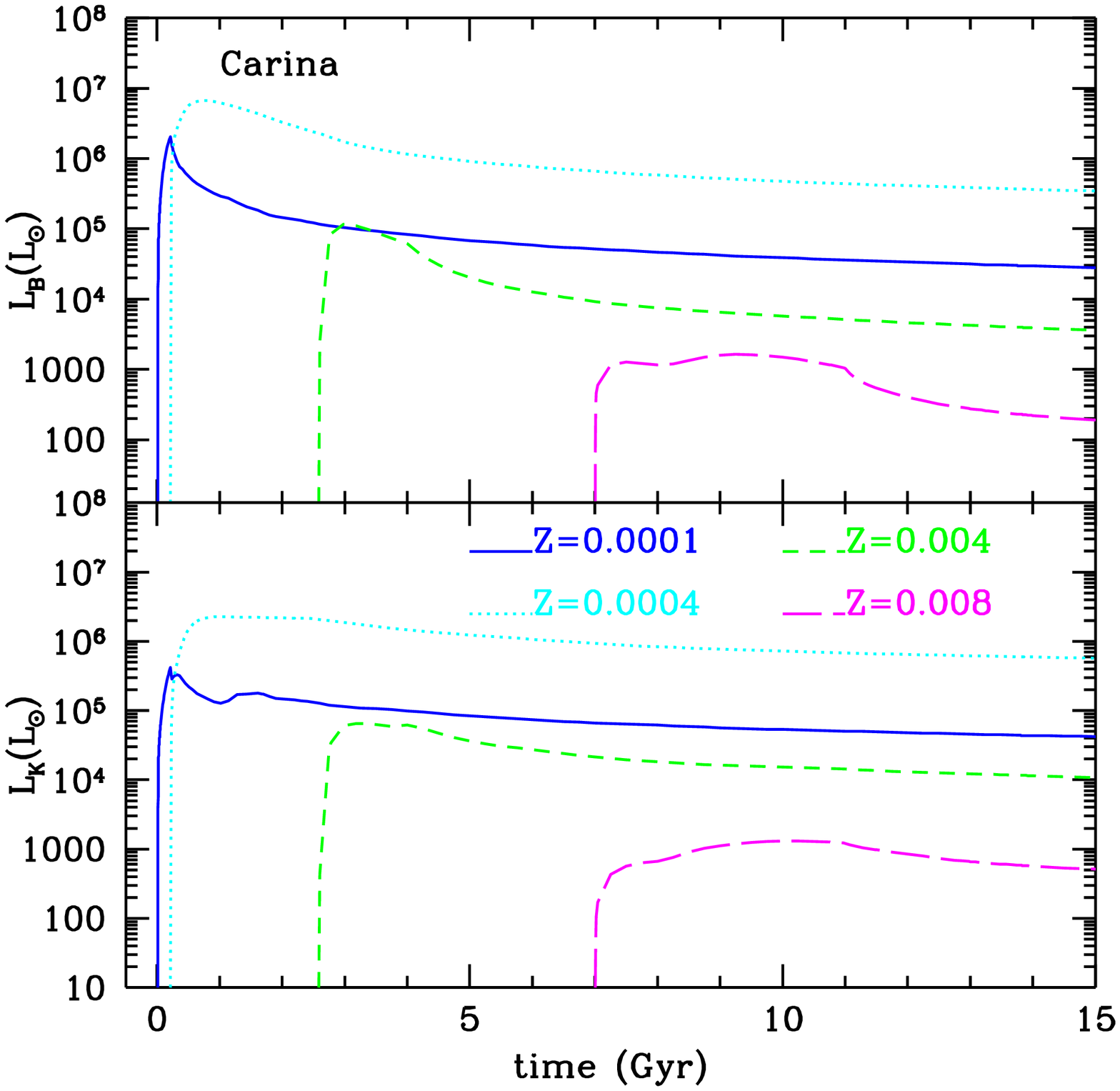}
\caption{ Predicted B- (upper panel) and K- (lower panel) band luminosities 
for the different stellar populations of the Carina dSph formed
at various metallicities as a function of time. }
\label{carina}
\end{figure}

\begin{figure}
\centering
\vspace{0.001cm}
\includegraphics[height=18pc,width=18pc]{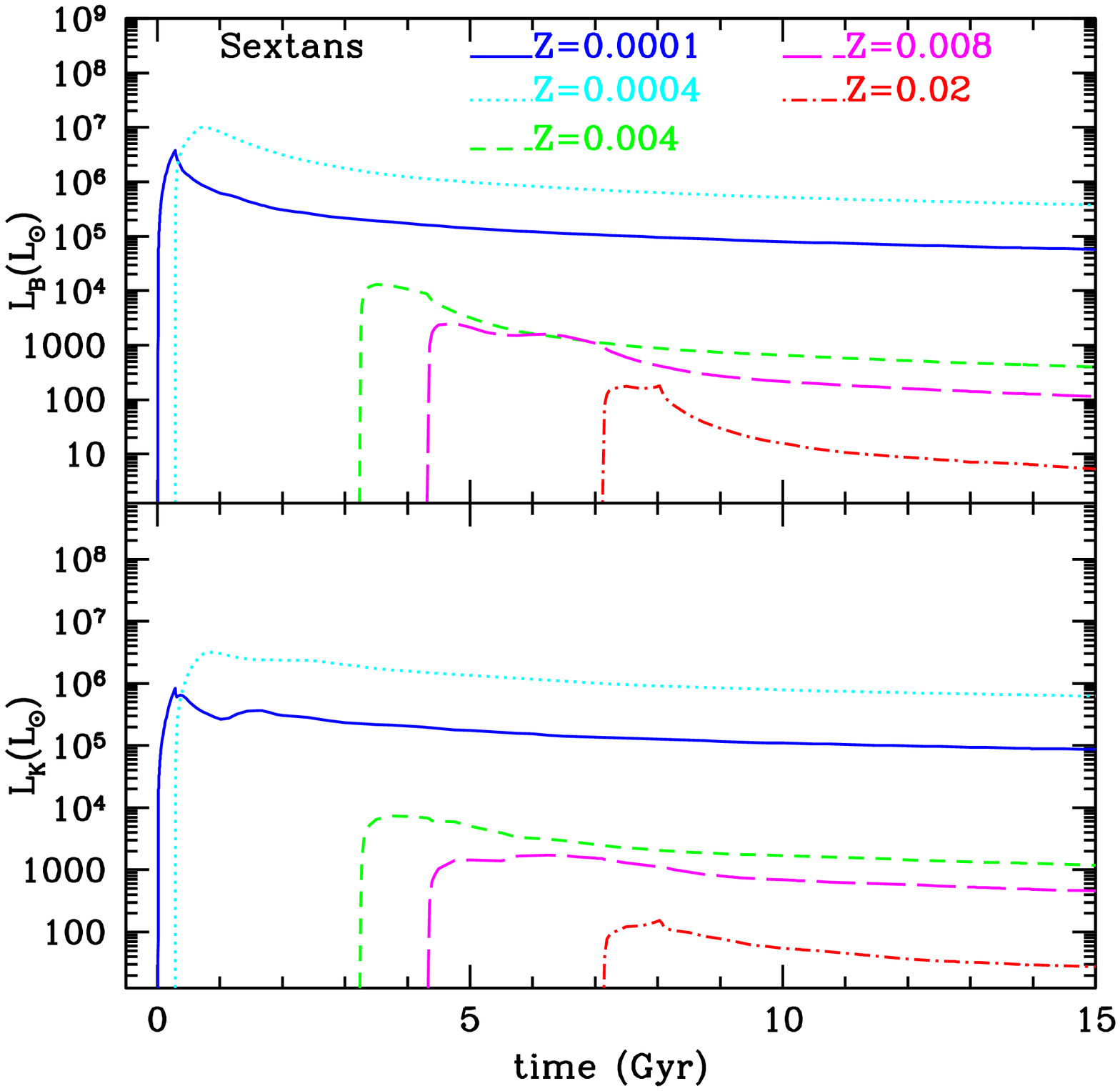}
\caption{ Predicted B- (upper panel) and K- (lower panel) band luminosities 
for the different stellar populations of the Sextans dSph formed 
at various metallicities as a function of time. }
\label{sextans}
\end{figure}

\begin{figure}
\centering
\vspace{0.001cm}
\includegraphics[height=18pc,width=18pc]{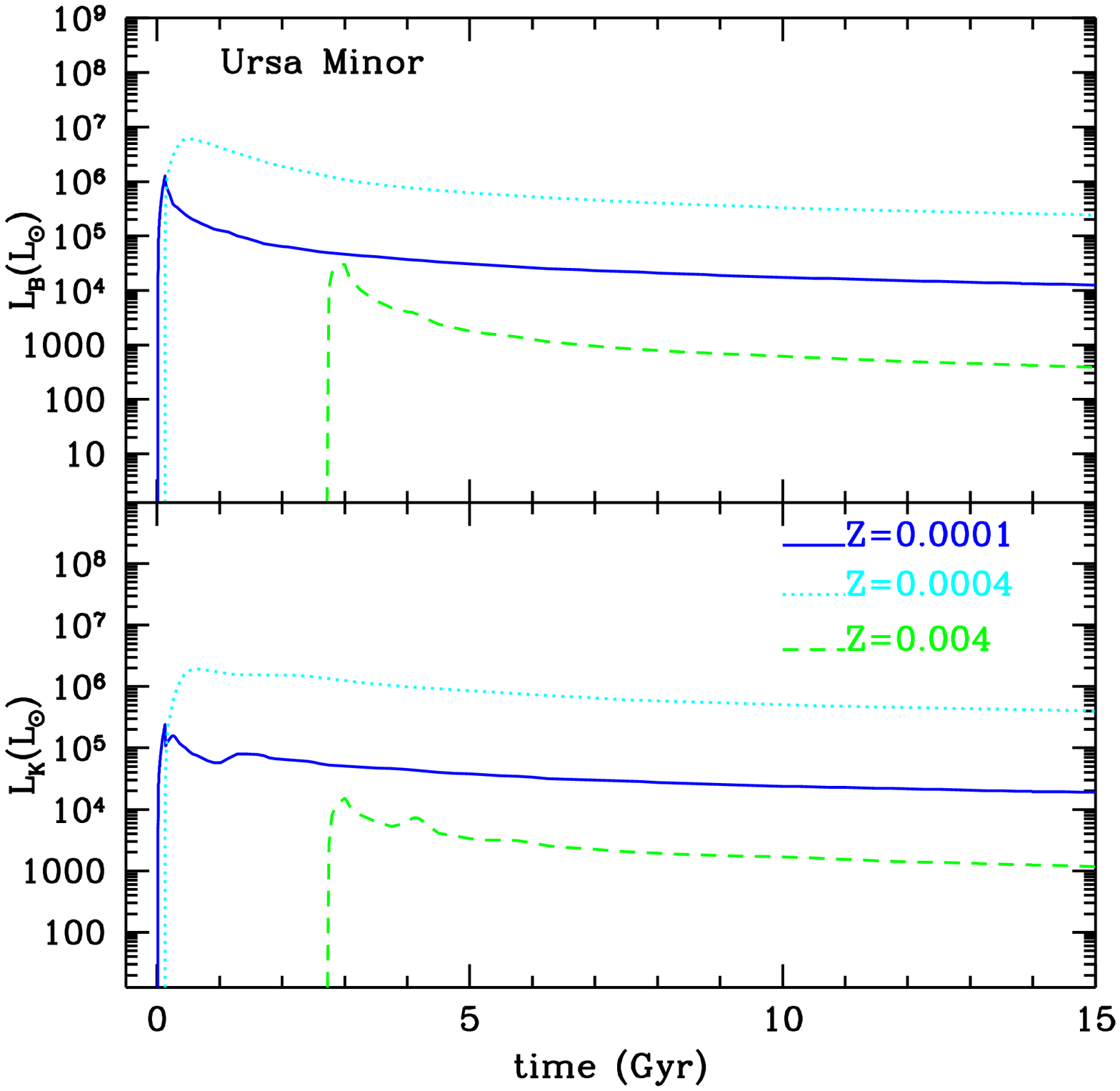}
\caption{ Predicted B- (upper panel) and K- (lower panel) band luminosities 
for the different stellar populations of the Ursa Minor dSph formed
at various metallicities as a function of time. }
\label{uminor}
\end{figure}

\begin{figure}
\centering
\vspace{0.001cm}
\includegraphics[height=18pc,width=18pc]{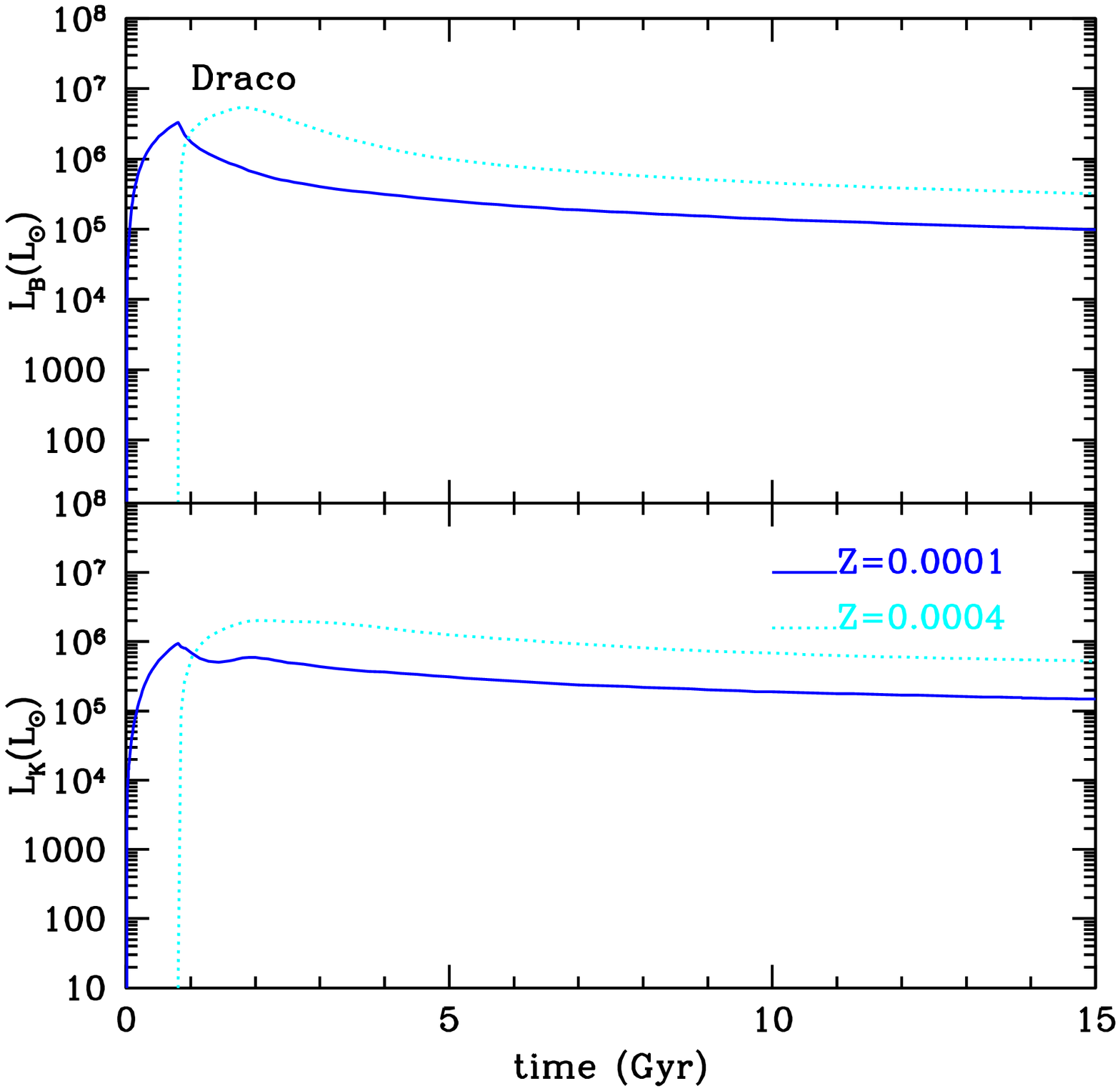}
\caption{ Predicted B- (upper panel) and K- (lower panel) band luminosities 
for the different stellar populations of the Draco  dSph formed
at various metallicities as a function of time. }
\label{draco}
\end{figure}

\begin{figure}
\centering
\vspace{0.001cm}
\includegraphics[height=18pc,width=18pc]{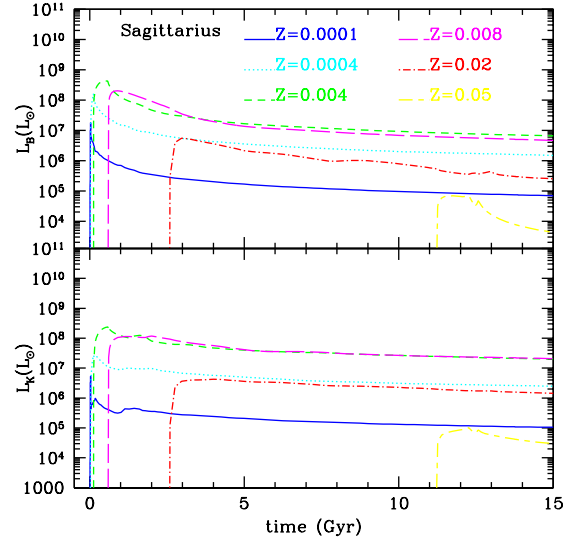}
\caption{ Predicted B- (upper panel) and K- (lower panel) band luminosities 
for the different stellar populations of the Sagittarius  dSph formed
at various metallicities as a function of time. }
\label{sagittarius}
\end{figure}

\subsection{The evolution of the V band luminosities}
Each dSph galaxy is assumed to be composed of different stellar populations, each one characterised by a particular 
metallicity. The metallicity values for the stellar populations are the ones of the BC spectro-photometric model, 
i.e.  $Z= 0.0001, 0.0004, 0.004, 0.008, 0.02 (Z_{\odot}), 0.05$.  
The luminosity of a dSph galaxy is given by the sum of the luminosities of the different SSPs. 

In
figures~\ref{sculptor},~\ref{carina},~\ref{sextans},~\ref{uminor},~\ref{draco},~\ref{sagittarius}
we show the predicted evolution of the luminosities for the different
stellar populations,   formed at different $Z$,  composing the
Sculptor, Carina, Sextans, Ursa Minor, Draco and Sagittarius dSphs,
respectively.  In most of the cases, the total luminosities  of the
dSph galaxies are dominated by the SSPs formed at the  lowest
metallicities, $Z=0.0001$ and $Z=0.0004$.  This is true for the
Sculptor, Carina, Sextans, Ursa Minor and Draco dSphs.  
For each galaxy, the stellar population dominating the metallicity distribution (LM04) 
dominates also the total K and B light. 
For all of
these galaxies, the stellar populations with metallicities $Z>0.0004$
give negligible contributions to the  total luminosities. A particular
case is represented by the Sagittarius dSphs.  In this galaxy, the B
and K luminosities are dominated by stellar populations with higher
metallicities, i.e. $Z=0.004$ and  $Z=0.008$. This is reflecting the fact 
that the Sagittarius galaxy had a star formation  history 
substantially different from the other 5 dSphs studied in this paper (see Fig.~\ref{sfr_gas}).
From the chemical evolution point of view, the difference
between Sagittarius and the other dSph galaxies studied here resides in the
fact that the former is characterised by a star formation efficiency
higher ($>$ 10 times) than the others. The slow evolution and the
occurrence of an intense galactic wind, which prevents metal rich
stars to be formed in a significant number, explains the fact that
stellar populations with low metallicities dominate the total
luminosity of these other dSph galaxies besides Sagittarius. 

\begin{figure}
\centering
\vspace{0.001cm}
\includegraphics[height=21pc,width=21pc]{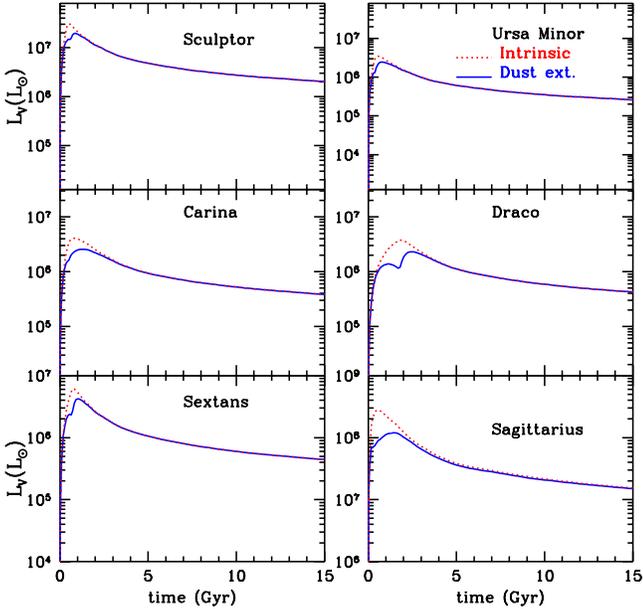}
\caption{ Predicted 
intrinsic (dotted lines) and dust obscured (solid lines) 
total V band luminosities 
for the six dSphs studied in this paper as functions of time. }
\label{Vband}
\end{figure}

The total luminosities of the dSph galaxies in the bluest bands of the optical spectra (i.e. U, B, V) 
are affected by dust extinction effects. These effects are negligible at the present time, but they must 
have played an important role in the past. 
In figure~\ref{Vband}, we show the predicted evolution of the V band Luminosities for the six dSph galaxies, 
both intrinsic and attenuated by dust extinction. For all the galaxies, the effects of dust extinction are noticeable during 
the star formation phase, when the galaxies have rich reservoirs of gas. After the onset of the 
galactic winds and the consequent decrease of the star formation, for each galaxy 
almost all of the interstellar gas is swept away and the effects of extinction become negligible. 
At the present time (13 Gyr), the dSphs have very little gas and the intrinsic and attenuated V band 
luminosities cannot be distinguished.

\subsection{The evolution of the colours of dSphs}
\begin{figure}
\centering
\vspace{0.001cm}
\includegraphics[height=19pc,width=19pc]{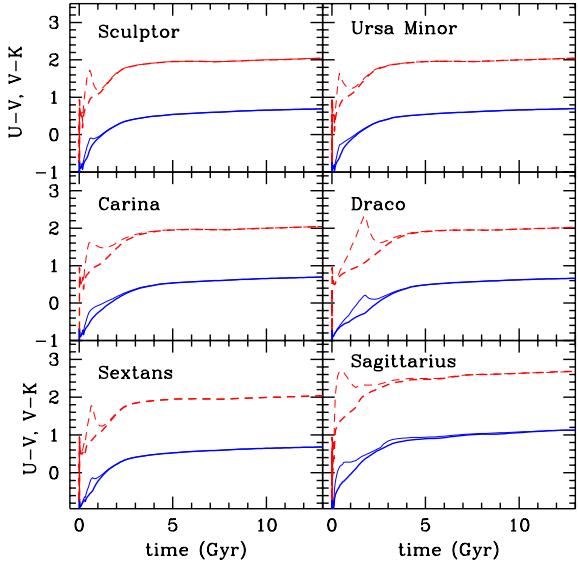}
\caption{ Predicted U-V (solid lines) and V-K (dashed lines) colours 
for the six dSphs studied in this paper as functions of time.  
The thick lines represent the intrinsic predicted colours.
The thin lines represent the colours calculated by 
taking into account dust extinction effects. 
}
\label{colours}
\end{figure}

In Figure~\ref{colours}, we show the predicted U-V (solid lines) and
V-K (dashed lines)  colours  for the six dSphs. 
It is
possible to see that during the star formation period, in general
lasting some Gyrs,  the attenuated colours are redder than the
intrinsic ones. After the end of the star formation, when the galaxies
have lost almost all of their gas, dust extinction is negligible and
the curves for the attenuated and the  intrinsic colours overlap.
This relatively short but important period corresponds to
the first few Gyrs of evolution, exactly the same epoch when the stars
formation is more intense. As mentioned in LMC06b, the evolution of
several dSphs is almost totally determined by the first few Gyrs after the beginning of the star formation. 
In this time interval, the pattern of the abundance ratios 
and the stellar metallicity distributions are almost totally in place 
(LMC06b), and also the colours are determined, supporting the idea
that the main activity in the dSph galaxies occurred very far in the
past and, after that, they evolved passively or with very low SF 
activity.\\ 
Finally, it is worth noting that the curve of the Draco dSph is substantially different from the others. 
This can be understood by looking at the star formation histories (Fig. ~\ref{sfr_gas}). The SFH of Draco peaks at 2 Gyr, corresponding to 
the age when the colour bump is largest in Fig. ~\ref{colours}.

\subsection{The colour-magnitude relation: a comparison with giant ellipticals}
\begin{figure}
\centering
\vspace{0.001cm}
\includegraphics[height=18pc,width=18pc]{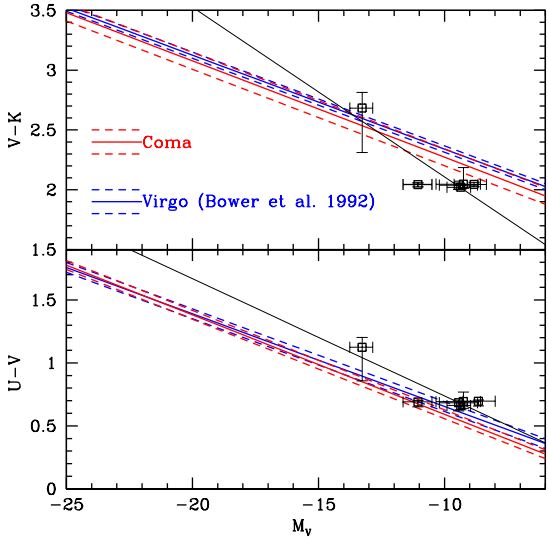}
\caption{ Predicted present-day U-V (lower panel) and V-K (upper panel) 
colours for the six dSphs studied in this paper (open squares with error bars). 
The blue and red solid thick lines represent the fits of the Bower et al. (1992) data for early type 
galaxies in the Virgo and Coma clusters, respectively,  
with their 1$\sigma$ scatter (dashed thick lines). The thin lines represent the best fit to the predicted colours, 
computed by means of a straight line $y=ax+b$. }
\label{cmr}
\end{figure}

In figure~\ref{cmr}, we plot the predicted present-day U-V (lower panel) and V-K (upper panel) 
colours as a function of the integrated absolute V magnitude for the six dSphs studied in this paper (open squares). 
Our predictions are  compared to the best fits (thick solid lines) of the colour-magnitude relations as 
observed by 
Bower at al. (1992), for the giant elliptical  galaxies in the Coma and Virgo clusters. 
The dashed lines represent the 1$\sigma$ scatter to the data by Bower et al. (1992). 
For the early-type galaxies in the Virgo and Coma clusters, 
Bower et al. (1992) found a small scatter of $\sim 0.05$ 
in the colours as a function of the total V magnitude. This small scatter implies a 
uniformity in the star formation histories of these systems, 
as well as a similar formation epoch (Renzini 2006).  By comparing the data by Bower et al. (1992) 
to the colour-magnitude relations (CMRs) predicted for the dSphs, it is possible to 
 see differences between
the dSphs and giant ellipticals found in clusters, in particular concerning their star formation history. 
Furthermore, it is possible to test whether dwarf local 
spheroidals belong to the same class of objects, i.e. if the dSphs represent the 
low-mass tail of the mass function of a single population of galactic spheroids.\\
The CMR for cluster galaxies by Bower at al. (1992) has been determined for galaxies 
with maximum apparent V magnitude 13 and $\sim 17$ in the case of the Virgo and Coma clusters, respectively. 
By adopting distance moduli of  31.1 and 35 for the Virgo (Freedman et al. 1994) and the 
Coma (Thomsen et al. 1997) clusters, respectively, the faintest absolute magnitudes 
which have been probed are $M_{V}\sim -18$, corresponding to galaxies considerably brighter than the local dSphs, presenting 
absolute magnitudes  $M_{V}\le -14$. 
For this reason, in Fig.~\ref{cmr}, we are comparing the  colours computed in this paper for the dSphs with the \emph{extrapolation} of the 
Bower at al. (1992) relation.
The most important aspect we are interested in is the slope of the CMR, since it is strictly linked to the galactic star formation history and 
age, in the sense that the tilt of the CMR tells us that fainter galaxies are bluer, hence have younger stellar populations. 
On the other hand, the scatter expresses how coeval the galaxies are in clusters and in the Local Group.\\
For each dSph, the colour has been computed using the 3 models introduced in Sect. 2.1, i.e. by means of the fiducial model, 
the minimal model,  and maximal model. This allows us to provide an error bar for the computed present-day colours of the six dSphs. 
By looking at the U-V vs V plot (lower panel), 
 we note that 5 out of 6 dSphs show very similar colours. 
This is due to the fact that for all the six systems but Sagittarius, the bulk of the star formation was completed 
already several Gyrs ago. Since 4 out 6 dSphs (Carina, Sextans, Ursa Minor and Draco) have very similar present-day luminosities 
and integrated absolute magnitudes (see fig. 8), they tend to cluster in the same region of the (U-V) vs V diagram. 
Sculptor is more luminous than these 4 dSphs, but presents similar present-day integrated colours. 
Sagittarius represents an exception, since it has a considerably higher luminosity and a redder colour. \\
In general, in the (U-V) vs $M_{V}$ plot, the dSphs are displaced to redder
colours than cluster dwarfs of comparable luminosity. However, 
The predicted (U-V) colours of the majority of the dSphs are compatible 
with the extrapolation to the faintest magnitudes of the 
colour-magnitude relation (CMR) for giant spheroids. 
We note that there is a partial overlap of the dSphs colours with the region including  the scatter of the Virgo and Coma CMRs. 
In particular, the error bars of 5 (Carina, Sextans, Ursa Minor, Sculptor and Sagittarius) 
out of the 6 systems analyzed in this work fall within the dispersion found by 
Bower et al. (1992).  
The other system lies outside of the dashed lines and presents a slightly ($\sim 0.1$ mag) redder (U-V) than 
the cluster dwarf ellipticals of the same magnitude. \\
The (V-K) colours of Carina, Sextans, Ursa Minor and Draco dSPhs are 0.2 mag  
bluer than the Coma spheroids of similar absolute magnitude, whereas Sculptor is 0.3 mag bluer. 
The (V-K) colour of Sagittarius, plotted with its error bar, is compatible with the Virgo and Coma CMRs. \\
We plot also the best fits to the predicted CMR of the dSphs, represented by a straight line $y=ax+b$.  
From Fig.~\ref{cmr}, it is clear that the slopes of he CMRs for dSphs and giant ellipticals are different. 
For the $(U-V)$ vs $M_{V}$ relation for dSphs, 
the slope of the line which best fits our predicted values is $a_{dSph}=-0.093$, against the values $a_{E,V}=-0.074 \pm 0.01$ and 
 $a_{E,C}=-0.079  \pm 0.007$ obtained by Bower at al. (1992) for the Virgo and Coma clusters, respectively. \\
From the study of the $(V-K)$ vs $M_{V}$ plot, we find that dSphs have an even steeper CMR, with  $a_{dSph}=-0.14$, against 
$a_{E,V}-0.078 \pm 0.007 $ and $a_{E,C}-0.08 \pm 0.013$. 
The differences between the CMRs of giant ellipticals and dSphs, in particular concerning the (V-K) vs V plot,  are very 
likely due to a more protracted SFHs  of the dSphs with respect to the ones of the cluster ellipticals. 
In fact, the CMDs of dSphs indicate that 
in general, they have experienced long star formation histories, 
lasting several Gyrs (see Fig. 1, Aparicio et al. 2001, Rizzi et al. 2003), at variance with their giant counterparts, 
which formed the bulk of their stars in 0.2-0.5 Gyr. 
In the case of the giant ellipticals, 
the scatter of the CMR as derived by Bower et al. (1992) implies  a dispersion in the age of the spheroid stellar 
populations of $\Delta_{form}\sim 2 $ Gyr.   
Although presenting similar present-day colours, 
the dSphs present a larger dispersion in the age of their stellar populations: for instance, the SFH of the Ursa Minor 
dSph is completed after 3 Gyrs, whereas Carina undergoes star formation episodes until 10 Gyrs of evolution (see Fig. 1). 
This implies that the formation of dSphs occurred with a lower degree of synchronicity 
than the formation of cluster ellipticals. 
Although presenting a larger dispersion in the ages, in the colour-magnitude plots it is not unlikely 
that the local dSphs  studied here form a sequence similar to the one observed for their giant 
 counterparts in clusters. 
The apparent differences between the colours of local dSphs and cluster dwarfs may be due to environmental effects: 
the clusters are in general denser environments, where interactions with the intra-cluster medium or with other galaxies 
are more frequent than in the Local Group and may have some effects also on the photometric evolution of their members. 
By means of cosmological simulations, Bullock et al. (2001) showed that dark matter halos in denser environments form 
earlier than more isolated halos. For this reason, cluster dwarfs may form earlier or may evolve faster, showing at the present day 
redder colours than local dSphs.\\
Since the sample considered here is limited to six systems, from the current analysis 
it is not possible to draw firm conclusions on this issue, i.e. whether local dSphs may represent the low-mass tail of 
giant ellipticals.  
In the future, it will be important to extend our analysis to a larger number of dSphs, 
in order to improve the statistics, as well
as to extend observational surveys in clusters to fainter absolute magnitudes. 
In the next Section, we will discuss implications of our analysis for the formation of spheroids.

\subsection{Implications for the formation of the giant galaxies}

An important question concerns the possibility that dSphs could represent the building blocks of 
the local giant ellipticals. 
Chemical evolution studies indicate that the dSphs are unlikely to represent the building blocks of giant spiral galaxies. 
This is suggested  by 
the abundance patterns observed in most of the dSph stars, which are clearly distinct from the ones observed in the 
stars of the Milky Way halo and disc (Venn et al. 2004). 
This fact represents a challenge to all 
galaxy formation models  
based on the popular $\Lambda$ cold dark matter ($\Lambda$CDM) cosmological scenario.
According to these models, 
in a $\Lambda$CDM-dominated   
universe, small dark matter halos are the first to collapse, then interact and merge 
to form larger halos. 
In this framework, massive spheroids are formed from 
several merging episodes among smaller objects (Somerville et al. 2001, Menci et al. 2002). 
This scenario for the most massive galaxies is at variance with the 
old monolithic collapse scenario, where ellipticals and bulges formed at high redshift (e.g. $z > 2-3$) as the
result of a violent burst of  star formation following a ``monolithic
collapse'' of a gas cloud (Larson 1974, Sandage 1986,  
Arimoto \& Yoshii 1987, Matteucci 1994).  \\
A picture where small and large spheroids form quasi-monolithically, although with different star formation timescales, 
 accounts naturally for the fact that local dwarf spheroidals may 
represent the low mass counterparts of giant 
ellipticals. A possible link between dSphs and giant ellipticals was suggested by Tamura, Hirashita \& Takeuchi (2001), 
who showed that both the mass-metallicity and mass-luminosity relations of some dSph galaxies 
may be understood as a low-mass extension of giant elliptical galaxies.\\
The link between dSphs and giant ellipticals is supported by 
the scenario proposed by Zaritsky et al. (2006), who found that small and giant 
spheroids belong to the same family and that a ``fundamental manifold'' for spheroids of all masses exist, i.e. 
that the local spheroids of all masses follow the same fundamental plane, with very few exceptions.  
As argued also by Zaritsky et al. (2006), it is probably difficult to account for the tightness 
of the CMR and of the spheroid manifold in a hierarchical framework, where the dSphs are the few survivors or 
intense merging activity.  In the hierarchical paradigm, the dwarf spheroidal galaxies should be the first 
galaxies to form their stars. The giant galaxies should assemble later by progressive merging of smaller systems. 
As a consequence, in general the dwarf galaxies should show stellar populations redder than their giant counterparts. 
A scenario in which dSphs were the building blocks of giant ellipticals would give rise to a CMR 
for dwarf and giant spheroids characterised by a shallower slope 
and a larger dispersion than the observations indicate, as shown by Kauffmann \& Charlot (1998) by means of semi-analythical models 
and by Saro et al. (2006) by means of numerical simulations. These papers showed that, in a scenario where the galaxies grow hierarchically, 
the present-day tilt and tightness of the CMR represents one of the most challenging features to explain. \\
Other open questions in cosmological 
studies of dSphs concern the ``missing satellite'' problem, i.e. the discrepancy between the observed number of 
dwarf galaxies clustered around giant spirals and the numbers predicted by hierarchical models (Moore et al. 1999) 
and their anisotropical spatial distribution (Kroupa et al. 2005), difficult to account for by current $\Lambda$CDM models. 
Together with the arguments studied in this section, 
these points outline our incomplete understanding of the process of galaxy formation. 
\renewcommand{\baselinestretch}{1.0}
\begin{table*}
\centering
\caption{Photometric properties of Local Group Dwarf galaxies. In column 1, we list all the six dwarf spheroidal galaxies.  
In columns 2 and 3, we present the observed and predicted present-day integrated (B-V) colours, respectively. 
In columns 4 and 5, we present the observed and predicted present-day HI 
to blue luminosity ratios (M$_{HI}$/L$_{B}$), respectively. }
\begin{tabular}{lcccccccc}
\\[-2.0ex] 
\hline
\\[-2.5ex]
\multicolumn{1}{l}{Galaxy}&\multicolumn{2}{c}{$(B-V)$}&\multicolumn{1}{c}{}&\multicolumn{2}{c}{M$_{HI}$/L$_{B}$}\\
\hline
\multicolumn{1}{c}{}&\multicolumn{1}{c}{Obs}&\multicolumn{1}{c}{Pred}&\multicolumn{1}{c}{}&\multicolumn{1}{c}{Obs}&\multicolumn{1}{c}{Pred}&\multicolumn{1}{c}{}\\
\hline 
\hline
\\[-1.0ex]
Sculptor    & 0.7$^1$             &  $0.67 \pm 0.02$          &   &    0.01$^1$           & 0.0006-0.0018                 \\
Carina      & 0.7$^1$             &  $0.67 \pm 0.05$          &   &    $<$0.002$^1$       & 0.008-0.022          \\
Sextans     & 0.7$^1$             &  $0.66 \pm 0.04$          &   &    $<$0.001$^1$       & 0.001-0.005                     \\
Ursa Minor  & $1.3 \pm 0.3$$^1$   &  $0.67 \pm 0.02$          &   &    $<$0.002$^1$       & 0.007-0.017                      \\
Draco       & $0.95 \pm 0.2$$^1$;$0.5 \pm 0.7$$^2$  &  $0.65 \pm 0.02$ &   &    $<$0.007$^{1, 2}$  & 0.008-0.036                     \\
Sagittarius & 0.6$^1$             &   $0.8 \pm 0.01$  &           &    $<$0.001$^1$       & 0.0012-0.0016                      \\
\hline
\hline
\end{tabular}
\flushleft
References: $^1$Mateo 1998, $^2$Aparicio et al. 2001\\
\end{table*}

In Table 3, we present a summary of some of  the observed and predicted 
photometric properties of the dSph galaxies studied in this work. 
In column 1, we list all the six dwarf spheroidal galaxies.  
In columns 2-5 we present the observed and predicted $(B-V)$ and 
HI to blue luminosity ratios (M$_{HI}$/L$_{B}$)  for the six 
dSphs, respectively. 
For the present-day $(B-V)$ values of Sculptor, Carina, Sextans, Ursa Minor and  Sagittarius we use the observational data published by Mateo (1998). 
Some of these data are presented without error bars. For Draco dSph, we also use the observational B, V magnitudes 
and M$_{HI}$/L$_{B}$ as published  by Aparicio et al. (2001).

By considering the error of 0.2-0.3 given by Mateo (1998) for the (B-V) colours of Ursa Minor and Draco as representative for 
all the other systems, we note that the B and V magnitudes of all the galaxies considered here are reproduced with good accuracy, with one exception. 
According to the data by Mateo (1998), the Ursa Minor dSph is considerably redder than what we predict, with a  discrepancy 
between our predictions and the observed $(B-V)$ of 0.6 mag.  
Interestingly, the observed $(B-V)$ is considerably redder even than giant ellipticals in local clusters. In fact, in the Fornax 
cluster, Karick et al. (2003) observe for elliptical galaxies $(B-V)$ ranging from 0.6 mag up to 1.1 mag.  
This tells us that the star formation history of the Ursa Minor was probably more complex than the one modelled in this paper. 
Moreover, a recent 
large-area photometric survey of the Ursa Minor dSph showed a peculiar structure of this galaxy, 
with an elongated morphology possibly due 
to tidal interaction with the Milky Way (Bellazzini etal. 2002). 
In the light of these results, Bellazzini et al. (2002) recomputed the integrated V magnitude 
of Ursa Minor finding $M_{V}=-10.3 \pm 0.4$, against the value $M_{V}=-8.9$ computed by Mateo (1998) without taking account any distortion effect. 
Unfortunately, no up-to-date estimations are available for the corrected integrated B magnitude. We await 
further data able to confirm or to rule out the peculiar (B-V) colour of the Ursa Minor dSph.\\

As shown in Tab. 3, in most of the cases, a scenario like ours, 
where the bulk of the dSphs stars formed prior to any interactions among them and with their 
largest conterparts, is able to account for the observed integrated colours of the local dSphs. 
Throughout this paper, we have assumed that the interactions between the dSphs and the external environment and with 
the giant galaxies of the Local Group 
had  negligible effects on their stellar populations, in the sense that they must have occurred after the bulk of the stars were formed. 
However, it is important to note that effects such as tidal interactions with the Milky Way 
may play some role in the evolution of dSphs.  The most striking example 
is represented by the Sagittarius dSph, which is characterised by 
a tidal tail extending from its centre out to a distance of $\sim 20 - 40$ kpc (Mateo et al. 1996, Majewski et al. 2003). 
In this case, such an interaction may have triggered a star formation episode a 
few Gyrs ago, as indicated by the M giants observed in its stream. 
Another case is represented by Carina. For this galaxy, photometric studies of regions beyond the classical tidal radius 
show a stellar distribution extending as far as several degrees from its centre (Kuhn et al. 1996, Mu\~noz et al. 2006)
All of these results indicate that the main effects of these interactions affected the last few Gyrs of evolution and 
were mainly of a dynamical nature (see Klimentowski et al. 2006), 
without affecting significantly the past star formation history of the dSphs. 

On the other hand, for most of the dSphs we overestimate the observed (M$_{HI}$/L$_{B}$), with the exception of the Sculptor dSph. 
This is again  due to the fact that our models 
do not include interactions with the external environment, in this case 
dynamical gas removal mechanisms, such as ram pressure stripping or tidal interactions. These dynamical processes, 
beside galactic winds, are likely to 
play a non negligible role in sweeping the gas from the dSph galaxies (Grebel et al. 2003). To investigate the 
effects of ram pressure stripping, hydrodynamical simulation studies should be required (see Marcolini et al. 2006).\\
In the case of the Sculptor dSph, we severely underestimate the observed  (M$_{HI}$/L$_{B}$) 
reported by Mateo (1998). Observations of this galaxy indicate that it  presents a particularly high content of HI. 
A possible explanation is that the Sculptor dSph is rapidly accreting gas from the external clouds, as 
its proximity to high velocity clouds is most likely suggesting (Carignan et al. 1998, 1999).


\begin{figure}
\centering
\vspace{0.001cm}
\includegraphics[height=18pc,width=18pc]{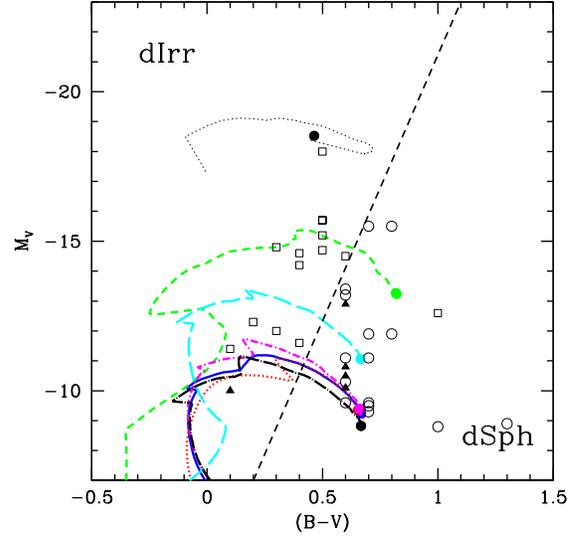}
\caption{ Observed and predicted V magnitude vs (B-V) colours for various dwarf galaxies. 
The thick solid lines represent the evolution of the V magnitudes as a function of the (B-V) colour
for the six dSphs studied in this work 
(solid line: Carina; dotted line: Draco; 
short-dashed line: Sagittarius; long-dashed line: Sculptor; 
dot-short-dashed: Sextans; dot-long-dashed: Ursa Minor). 
The solid dots at the end of each line represent the present-day values. 
The dotted thin line represents the predictions obtained by means of 
a model for the Large Magellanic cloud.  
The open circles, open squares and solid triangles are the observed values for dSph galaxies, dIrrs and 
transition objects, respectively (data from Mateo 1998). The dashed line 
is the same as the one drawn in figure 5 of Mateo (1998) and separates the systems classified as dIrrs (left region of the plot) 
and dSphs (right region). 
}
\label{BV}
\end{figure}

\begin{figure}
\centering
\vspace{0.001cm}
\includegraphics[height=18pc,width=18pc]{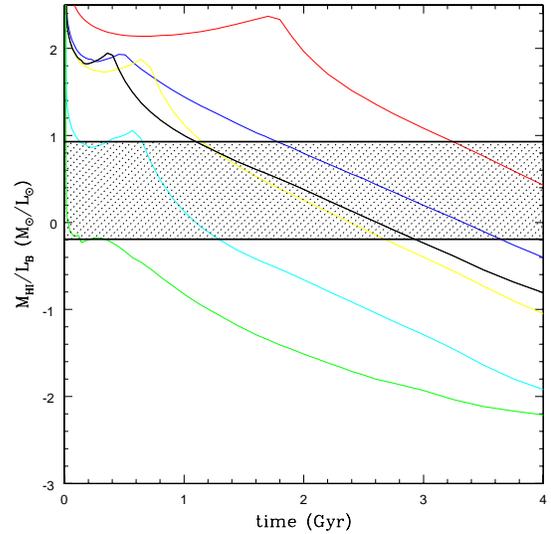}
\caption{ Evolution of $M_{HI}/L_{B}$ four the six dSphs (solid lines) during their initial 4 Gyr of evolution, 
corresponding approximately to the time that each dSph galaxy spends in the left part of the plot in Fig.~\ref{BV}.
The shaded area encloses all the possible values for  $M_{HI}/L_{B}$ observed in local dIrr (Mateo 1998). }
\label{mhi}
\end{figure}


\subsection{What do the colours tell us of the evolutionary history of dSph galaxies?} 
An interesting question concerning dSphs is how their SFHs
have proceeded in the past. To investigate this aspect, of particular interest is the study of the relationship 
between dSphs and dwarf irregulars which  could represent the same objects seen in different evolutionary phases. 
In general, dSphs have physical properties rather different than dIrrs 
(Grebel et al. 2003 and references therein): (1) the gas content, with dSphs being gas-poor, whereas dIrrs contain large reservoirs of gas;  
(2) Rotation: dSphs have little angular momentum, 
whereas dIrrs are fast rotating systems; (3) Metallicity: dSphs present 
 metallicities  in general higher than dIrrs at the same optical luminosity. All these arguments stand against a continuous  
evolution from dIrrs to dSphs and seem to indicate that they represent systems with different origin and 
past SFHs. \\

LM03 analyzed a possible connection between the dSph galaxies and Blue Compact Dwarf Galaxies (BCGs). 
The SFHs inferred from the CMDs of BCGs are characterised by short starbursts separated by periods of inactivity 
(Schulte-Ladbeck et al. 2001), similar to the ones derived from the CMDs of dIrr galaxies (Kunth \& \"Ostlin 2000, Annibali et al. 2003). 
LM03 claimed that the chemical evolution of BCGs and dSphs indicates  marked 
differences in their SFHs. 
While the observed
distributions of several abundance ratios in BCGs are reproduced by
models with SF characterised by short and repeated bursts separated by
long quiescent periods, the  abundance ratios of local dSph galaxies
require a SF which  proceeds in one long episode of low activity. 
The SF efficiencies of the two models are basically different, as well 
as the efficiencies of the galactic winds, 
which are much higher ($w_i$ = 4 - 13) for the dSphs than in the case of the BCGs  ($w_i$ = 0.25 - 1). 
They also concluded that a bursting model, suited to describe the SFHs of both dIrrs and BCGs, 
is not able to reproduce the dSphs data and vice-versa. \\
However, some other evidence is in  favour of a possible transition from dIrrs to dSphs, in particular,  
(a) the similarity of the exponential radial brightness profiles and, (b) the existence of the ``transition type'' 
dwarf galaxies, i.e. systems presenting properties common to both kinds of objects, such as 
old stellar populations (typical of dSphs) and high gas content (typical of dIrrs). 
Some important indications of the past evolution of dwarf galaxies come from the study 
of their integrated absolute V magnitude as a function of their (B-V) colour (Mateo 1998). 
In Figure~\ref{BV}, we plot the predicted evolution of the  integrated V magnitude  as a function of the (B-V) colour for the 
six dwarf spheroidal galaxies studied in this work 
(solid thick lines). The dotted thin line indicates the evolution of the $M_{V} $ vs (B-V) for a model representing the Large Magellanic Cloud (LMC). 
This model is the same as the one used by Calura, Matteucci \& Vladilo (2003) to study the chemical evolution of the Damped Lyman Alpha absorbers.  
For further details on this model, we refer the reader to that paper. This model is used to represent the evolution of the V magnitude 
as a function of the (B-V) colour for a typical dIrr galaxy. 
The solid circles at the end of each line represent the present-day values. 

Along with the predicted values, in Fig.~\ref{BV} we plot also the observed values for three different kinds of objects belonging to the Local Group (Mateo 1998):  
the dSph galaxies (open circles), the dIrrs (open squares) and the  transition objects (solid triangles). The dashed line 
is the same as the one drawn in figure 5 of Mateo (1998) and it is helpful to show how, in this plot, 
the dIrrs tend to be segregated from the dSphs. The transition objects are in general located close to the dashed line. 
It is interesting to see how, in general, at the same V magnitude, dIrrs tend to be bluer than dSph galaxies. 
Our predictions indicate that, in their past history, when they were bluer and more gas-rich than today,
the dSph galaxies used to lay on the left of the dashed line, in the region where the present day dIrrs  lie.
This is visible in the left part of the plot, where the solid thick lines overlap with the open squares. 
According to our results, during their past history, the dSphs 
may have shown photometric properties similar to those of the present day dIrrs. 

Our results show that dSphs and dIrrs may share a common progenitor phase, as pointed out also by 
Binggeli (1994) and Skillmann \& Bender (1995). 
During this phase, both dSphs and dIrrs may have occupied the left region of V vs (B-V) plot and then  
they may  have evolved through different branches.  
Since then, dIrrs have experienced little evolution and they have not reached the right side of the $M_{V}$ vs (B-V) plot. 
This seems to be suggested also by the predicted evolution of V vs (B-V) for the LMC \footnote{ Notwithstanding the large mass of the LMC, 
which renders it probably too massive to be regarded as an average dIrr galaxy, its evolution in the $M_{V}$ vs (B-V) plot shows that 
it never crossed the dashed line. For this reason, from the photometric point of view, in this plot 
it can be regarded as an example of a typical dIrr. }
  galaxy in Fig. 10. 

On the other hand, 
dSphs have experienced a stronger evolution and they have moved to the right part of the V vs (B-V) plot. 
It is nevertheless possible that a sub-class of dSphs may derive from dIrrs. 
The transition galaxies may represent this sub-class of objects, which are presumably younger than the dSphs. \\
Finally, it is interesting to see whether in the past, the gas content of dSphs was compatible with what local observations of dIrrs indicate. 
In Fig. ~\ref{mhi}, we plot the time evolution of $M_{HI}/L_{B}$ for the six dSphs during their initial 4 Gyr of evolution, 
corresponding approximately to the time that each dSph galaxy spends in the left part of the plot in Fig.~\ref{BV}, which can be regarded 
as the ``dIrr'' phase. The shaded area encloses all the possible values for  $M_{HI}/L_{B}$ observed in local dIrr as published by Mateo (1998). 
We see that, for all dSphs with the exception of one case (Sagittarius),  
the phase when $M_{HI}/L_{B}>0.64 M_{\odot}/L_{\odot}$ (corresponding to the lowest  $M_{HI}/L_{B}$ observed in dIrrs, Mateo 1998) 
lasts more than 1 Gyr. In Fig.~\ref{mhi}, the region bracketed by the shaded area is only indicative of the past $M_{HI}/L_{B}$ values for dIrrs, 
since in the past dIrrs are likely to have shown $M_{HI}/L_{B}$ values larger than the current ones.  
This confirms that during their first few Gyrs of evolution, the majority of dSphs would 
have looked like the dIrr as we see them today. \\

\section{Summary and Conclusions}

By means of a chemical evolution model combined with a
spectro-photometric code we were able to predict the evolution of
several photometric properties of six Local Group Dwarf Spheroidal
galaxies (Carina, Draco, Sagittarius, Sculptor, Sextans and Ursa
Minor). The chemical evolution models adopt up-to-date
nucleosynthesis from intermediate mass stars, massive stars and SNe type II
and Ia and take into account the contribution of SNe to the
energetics of the ISM. 
For the six dSphs, the star formation histories 
are taken from the  observed colour-magnitude diagrams (CMDs, Dolphin et al. 2005). 
The proposed 
scenario for the evolution of these galaxies is characterised by low
star formation rates and high galactic wind efficiencies. Such a 
scenario allows us to predict colours and magnitudes in agreement with
observations.

The main conclusions can be summarized as follows: 

\begin{itemize}

\item
the total luminosities of 5 (Sculptor, Carina, Sextans, Ursa Minor and
Draco) out of 6 dSphs galaxies analysed here are dominated by the SSPs
formed at the lowest metallicities, i.e. $Z\le 0.0004$. This is a
consequence of the low SF efficiency and of the intense wind which
removes a large fraction of the gas of the galaxy, almost halting the
SF and preventing metal-rich stars to be formed. For each galaxy, the
stellar populations dominating the metallicity distribution (LM04)
dominate also the total light.

\item
in the case of Sagittarius, the B and K luminosities are dominated by stellar 
populations with higher metallicities, i.e. $Z\ge 0.004$,  due to the much 
higher SF efficiency adopted for this galaxy ($>$ 10 times higher than the ones of the other dSphs);

\item 
only during the SF phase, when all galaxies exhibit a high gas
content, the effects of dust extinction in the total luminosities in
the U, V, B bands are noticeable. After the onset of the galactic winds
and the consequent removal of the interstellar gas, the effects of
extinction become negligible;

\item
We compared 
the predicted current integrated (U-V) and (V-K) colours as a function of the absolute V magnitude for
the dSphs to the colour-magnitude relation for giant spheroids in clusters. 
The main differences between dSphs and the giant ellipticals concern their star formation histories. 
In particular, the formation of dSphs occurred with a lower degree of synchronicity 
than the formation of giant ellipticals.  
The differences may also be due to environmental effects, since clusters are environments denser than the Local Group.  
In the colour-magnitude plots, 
the six local dSphs  studied here seem to form a sequence similar to the one observed for their 
 counterparts in clusters, although with a slightly steeper slope and a larger dispersion in their ages.   
Although we can not draw any firm conclusion on this aspect, 
it is not unlikely that the dSphs can be regarded 
as the lowest-mass tails of their larger counterparts and not
the building blocks from which they were assembled. This issue represents a 
challenge to all galaxy formation models based on the popular 
$\Lambda$ cold dark matter cosmological scenario. 

\item
The study of the evolution of the integrated V magnitude as a function of the
(B-V) colour suggests that, during their past history, the dSphs
may have shown photometric properties similar to the ones of the
present-day dIrrs. 
It is possible that dSphs and dIrrs shared  a common progenitor phase, and then evolved 
through different paths. After the common phase, dIrrs have experienced little evolution, mantaining most of their gas 
and blue colours. dSphs have instead experienced a stronger evolution, 
losing all of their gas and showing present-time colours redder than dIrrs. Transition types might represent a sub-class 
of dIrrs, which may evolve into dSphs. 

\end{itemize}

\begin{acknowledgements}
We wish to thank Simone Recchi for a critical reading of the manuscript and for several useful comments and suggestions. 
We are grateful to an anonymous referee for several useful comments that improved the quality 
of our work. \\
G.A.L. thanks the hospitality of the Dipartimento di 
Astronomia-Universit\'a di Trieste which provided all the support
during his stay. 
G.A.L. acknowledges financial support from the Brazilian agency FAPESP (proj.
06/57824-1).
F.M. and F. C. acknowledge financial support  
from INAF (Italian National Institute of Astrophysics) project 2005\_1.06.08.16
\end{acknowledgements}


\begin{thebibliography}{}


\bibitem{}
Annibali F., Greggio L., Tosi M., Aloisi A., Leitherer C., 2003, AJ, 126, 2752

\bibitem{}
Aparicio A., Carrera R., Martinez-Delgado D., 2001, AJ, 122, 2524

\bibitem{}
Arimoto N., Yoshii Y., 1987, A\&A, 173, 23

\bibitem{}
Bellazzini M., Ferraro F. R., Origlia L., Pancino E., Monaco L., Oliva E., 2002, AJ, 124.

\bibitem{}
Binggeli B., 1994, in ``Panchromatic View of Galaxies'', eds. G. Hensler, Ch. Theis, J. Gallagher, 173

\bibitem{}
Bonifacio, P., Hill, V., Molaro, P. et al., 2000, A\&A, 359, 663 

\bibitem{}
Bonifacio, P., Sbordone, L., Marconi, G., Pasquini, L., \& Hill, V., 2004, A\&A, 414, 503

\bibitem{}
Bower R. G., Lucey J. R., Ellis R. S., 1992, MNRAS, 254, 613	

\bibitem{}
Bradamante F., Matteucci F., D'Ercole A., 1998, A\&A, 337, 338 

\bibitem{}
Bruzual,  A. G., Charlot, S., 2003, MNRAS, 344, 1000

\bibitem{}	
Bullock, J. S., Kolatt, T. S., Sigad, Y., Somerville, R. S., Kravtsov, A. V., Klypin, A. A., Primack, J. R., Dekel, A., MNRAS, 321, 559


\bibitem{}
Calura F., Matteucci F., Vladilo G., 2003, MNRAS, 340, 59

\bibitem{}
Calura F., Matteucci F., 2004, MNRAS, 350, 351

\bibitem{}
Calura F., Matteucci F., 2006, ApJ, 652, 889


\bibitem{}
Calura F., Matteucci F., Menci N., 2004, MNRAS, 353, 500       



\bibitem{}
Calzetti, D., Kinney, A. L., Storchi-Bergmann, T., 1994, ApJ, 429, 582

\bibitem{}
Calzetti, D., 1997, in ``The Ultraviolet Universe at Low and High Redshift : Probing the Progress of Galaxy Evolution'', William H. Waller et al. eds., AIP Conference Proceedings, 408, 403


\bibitem{}
Calzetti, D., 2001, PASP, 113, 1449 

\bibitem{}
Carignan, P., et al., 1998, AJ, 116, 1690

\bibitem{}
Carignan, P., 1999, PASA, 16, 18


\bibitem{}
Carrera R., Aparicio A., Martinez-Delgado D., Alonso-Garcia J., 2002, AJ, 123, 3199

\bibitem{}
Cescutti, G., Fran\c cois, P., Matteucci, F., Cayrel, R., \& Spite, M., 2006, A\&A, 448, 557

\bibitem{}
C\^ot\'e S., Freeman K. C., Carignan C., Quinn P. J., 1997, AJ, 114, 1313

\bibitem{}
Davies J.I., Phillips S., 1988, MNRAS, 233, 553

\bibitem{}
Dekel A., Silk J., 1986, ApJ, 303, 39

\bibitem{}
Dolphin A. E., 2002, MNRAS, 332, 91

\bibitem{} 
Dolphin, A.E., Weisz, D.R., Skillman, E.D., Holtzman, J.A., 2005, 
to appear in Valls-Gabuad D. \& Chavez M., eds.,
Resolved Stellar Populations, ASP Conference Series, astro-ph/0506430


\bibitem{}
Fenner Y., Gibson B. K., Gallino R., Lugaro M., 2006, ApJ, 646, 184

\bibitem{}
Ferguson H. C., Sandage A., 1991, AJ, 101, 765

\bibitem{}
Fran\c cois, P., Matteucci, F., Cayrel, R. et al., 2004, A\&A, 421, 613

\bibitem{}
Freedman, W. L., et al. 1994, Nature, 371, 757         


\bibitem{}
Ikuta, C., \& Arimoto, N., 2002, A\&A, 391, 55

\bibitem{}
Gallagher, J. S., Madsen, G. J., Reynolds, R. J., Grebel, E. K., Smecker-Hane, T. A., 2003, ApJ, 588, 326

\bibitem{}
Geisler, D., Smith, V.V., Wallerstein, G., Gonzalez, G., \& Charbonnel, C., 2005, AJ, 129, 1428

\bibitem{}
Geisler, D., Wallerstein, G., Smith, V. V., Casetti-Dinescu, D. I., 2007, 2007, PASP, 119, 939

\bibitem{}
Grebel E. K., 1998, in Andersen J., ed., Highlights of Astronomy, Kluwer, Dordrecht, 11, 125


\bibitem{}
Grebel E. K., 2001, in ``Dwarf Galaxies and Their Environment'', ed. K. S. De Boer, R. Dettmar, \& U. Klein (Aachen: Shaker Verlag), 45 

\bibitem{}
Grebel, E. K., Gallagher J. S., III, Harbeck D., 2003, AJ, 125, 1926


\bibitem{}
Greggio, L., \& Renzini, A., 1983, A\&A, 118, 217

\bibitem{}
Graham A., Guzm`an R., 2003, ApJ, 2936

\bibitem{}
Guzm`an R., Jangren A., Koo D.C., Bershady M.A., Simard L., 1998, ApJ, L13

\bibitem{}
Hernandez, X., Gilmore, G., \& Valls-Gabaud, D., 2000, MNRAS, 317, 831

\bibitem{}
Hurley-Keller D., Mateo M., Nemec J., 1998, AJ, 115, 1840

\bibitem{}
Irwin M., Hatzidimitriou D., 1995, MNRAS, 277, 1354

\bibitem{}
Karick, A. M., Drinkwater, M. J., Gregg, M. D.,  2003, MNRAS, 344, 188


\bibitem{}
Kauffmann G., Charlot S., 1998, MNRAS, 294, 705


\bibitem{}
Kawata D., Arimoto N., Cen R., \& Gibson B.K., 2006, ApJ, 641, 785


\bibitem{}
Klimentowski J., Lokas E. L., Kazantzidis S., Prada F., Mayer L., Mamon G. A., 2006, MNRAS, submitted, astro-ph/0611296

\bibitem{}
Koch A., Grebel E.K., Wyse R.F.G., et al., 2005, AJ, 131, 895

\bibitem{}
Koch A., Grebel E. K., Wyse R. F. G., Kleyna J. T., Wilkinson M. I., Harbeck D. R., Gilmore G. F., Evans N. W., 2006, AJ, 131, 895

\bibitem{}
Kroupa, P., Theis, C., Boily, C. M., 2005, A\&A, 431, 517

\bibitem{}
Kuhn J. R., Smith H. A., Hawley S. L., 1996, ApJ, 469, L93

\bibitem{}
Kunth D., \"Ostlin G., 2000, A\&AR, 10, 1 
 
\bibitem{}
Lanfranchi, G., \& Matteucci, F., 2003, MNRAS, 345, 71, LM03
 
\bibitem{}
Lanfranchi, G., \& Matteucci, F., 2004, MNRAS, 351, 1338, LM04
 
\bibitem{}
Lanfranchi, G., Matteucci, F., \& Cescutti, G., 2006a, MNRAS, 365, 477
 
\bibitem{}
Lanfranchi, G., Matteucci, F., \& Cescutti, G., 2006b, A\&A, 453, 67

\bibitem{}
Lanfranchi, G., Matteucci, F., 2007, A\&A,  468, 927, LM07

\bibitem{}
Larson R. B., 1974, MNRAS, 166, 585

\bibitem{}
Lin D.N.C., Faber S.M., 1983, ApJ, 266L, 21

\bibitem{}
Majewski S. R., Skrutskie M. F., Weinberg M. D., Ostheimer J. C., 2003, ApJ, 599, 1082


\bibitem{}
Marcolini, A., D'Ercole, A., Brighenti, F., Recchi, S., 2006, MNRAS, in press

\bibitem{}
Mateo M., Mirabal N., Udalski A., Szymanski M., Kaluzny J., Kubiak M., Krzeminski W., Stanek, K. Z., 1996, ApJ, 458, L13

\bibitem{}
Mateo M.L., 1998, ARA\&A, 36, 435
 
\bibitem{}
Matteucci, F., 1992, ApJ, 397, 32

\bibitem{}
Matteucci, F., 1994, A\&A, 288, 57

\bibitem{}
Matteucci, F., 1996, FCPh, 17, 283

\bibitem{} 
Matteucci F., Greggio L., 1986, A\&A, 154, 279

\bibitem{} 
Matteucci, \& F., Tornamb\'{e}, A., 1987, A\&A, 185, 51

\bibitem{} 
Mayer L., Mastropietro C., Wadsley J., Stadel J., Moore B., 2006, MNRAS, 369, 1021	

\bibitem{} 
Menci N., Cavaliere A., Fontana A., Giallongo E., Poli F., 2002, ApJ, 575, 18

\bibitem{} 
Moore B., Ghigna S., Governato F., Lake G., Quinn T., Stadel J., Tozzi P., 1999, ApJ, 524, L19

\bibitem{}
Monaco L., Bellazzini M., Bonifacio P. et al., 2005, A\&A, 441, 141

\bibitem{}
Mori M., Burkert A., 2000, ApJ, 538, 559

\bibitem{}
Mu\~noz R. R., et al., 2006, ApJ, 649, 201

\bibitem{} 
Nomoto K., Hashimoto M., Tsujimoto T., Thielemann F.-K., Kishimoto
 N., Kubo Y., Nakasato N., 1997, 
Nucl. Phys. A, 616, 79c

\bibitem{}
Papaderos P., Loose H.-H., Fricke K.J., Thuan T.X., 1996, A\&A, 314, 59

\bibitem{}
Phillipps S., Driver S. P., Couch W. J., Smith R. M., 1998 ApJ, 498, L119

\bibitem{}
Recchi S., Matteucci F., D'Ercole A., Tosi M., 2002, A\&A, 384, 799

\bibitem{}
Renzini A., 2006, ARA\&A, 44, 141

\bibitem{}
Ricotti M., Gnedin N. Y., 2005, ApJ, 629, 259
	
\bibitem{} 
Rizzi L., Held E.V., Bertelli G., Saviane I., 2003, ApJL, 589, 85

\bibitem{}
Sadakane K., Arimoto N., Ikuta C., Aoki W., Jablonka P., Tajitsu A., 2004, PASJ, 56, 1041

\bibitem{} 
Salpeter E.E., 1955, ApJ, 121, 161


\bibitem{} 
Sandage A., 1986, A\&A, 161, 89

\bibitem{} 
Saro A., Borgani S., Tornatore L., Dolag K., Murante G., Calura F., Biviano A., 2006, MNRAS, 373, 397

\bibitem{} 
Shapley H., 1938, Harvard Colleg Obs. Bulletin, 908, 1
\bibitem{} 
Smecker-Hane T. A., Stetson P. B., Hesser J. E., Vandenberg D. A., 1996, 
in ''From Stars to Galaxies: The Impact of Stellar Physics on Galaxy Evolution'', 
 ed. C. Leitherer, U. Fritze-von-Alvensleben, and J. Huchra, 
ASP Conference Series, 98, 328

\bibitem{} 
Schmidt M., 1963, ApJ, 137, 758

\bibitem{} 
Schulte-Ladbeck R. E., Hopp U., Greggio L., Crone M. M., Drozdovsky I. O., 2001, Ap\&SS, 277, 309

\bibitem{}
Shetrone M., C\^ot\'e P., Sargent W.L.W., 2001, ApJ, 548, 592
 
\bibitem{}
Shetrone M., Venn K.A., Tolstoy E., Primas F., 2003, AJ, 125, 684 

\bibitem{}
Skillman E. D., Bender R., 1995, RMxAC, 3, 25

\bibitem{}
Somerville R. S., Primack J. R., Faber S. M., 2001, MNRAS, 320, 504


\bibitem{}
Tamura, N.,  Hirashita, H., Takeuchi, T. T., 2001, ApJ, 552, L113


\bibitem{} 
Thielemann F.K., Nomoto K., Hashimoto M., 1996, ApJ, 460, 408

\bibitem{} 
Thomsen, Bjarne, Baum, William A., Hammergren, Mark, Worthey, Guy, 1997, ApJ, 483, L37


\bibitem{} 
Tinsley B.M., 1980, FCPh, 5, 287

\bibitem{}
Tolstoy E., Venn K.A., Shetrone M., Primas F., Hill V.,
Kaufer A., Szeifert T., 2003, AJ, 125, 707

\bibitem{}
Tolstoy, E., et al. 2004, ApJ, 617, L119 

\bibitem{}
van den Bergh S., 1994, ApJ, 428, 617

\bibitem{}
Van den Hoeck L.B., Groenwegen M.A.T., 1997, A\&AS,123 , 305

\bibitem{}
Venn K. A., Irwin M., Shetrone M.D. et al., 2004, AJ, 128, 1177

\bibitem{}
Woosley S.E., Weaver, T.A., 1995, ApJS, 101, 181 

\bibitem{}
Zaritsky D., Gonzalez A. H., Zabludoff A. I., 2006, ApJ, 642, L37

\end{thebibliography}
\end{document}